\documentclass[aps,prb,onecolumn,nofootinbib,citeautoscript,10pt]{revtex4-2}

\usepackage{amsmath,amssymb,bm} 
\usepackage{graphicx,comment}

\usepackage[tight]{subfigure} 

\usepackage[dvipsnames]{xcolor} 
\usepackage[papersize={8.5in,11in}]{geometry}
\usepackage[colorlinks=true]{hyperref}
\hypersetup{
    bookmarks=true,         
    unicode=false,          
    pdftoolbar=true,        
    pdfmenubar=true,        
    pdffitwindow=false,     
    pdfstartview={FitH},    
    pdfkeywords={keyword1} {key2} {key3}, 
    pdfnewwindow=true,      
    colorlinks=true,       
    linkcolor=magenta, 
    citecolor=blue,        
    filecolor=magenta,      
    urlcolor=blue           
} 

\usepackage{dcolumn}
\usepackage{color}
\usepackage{amssymb,amsmath, makecell}
\usepackage{tabularx, makecell}
\usepackage{latexsym}
\usepackage{colortbl}
\usepackage{psfrag}
\usepackage{bbm,bm,array,physics}
\usepackage{dsfont}
\usepackage{float, mathrsfs, upgreek}

\def \nn{\nonumber \\}

\begin{document}
\title{Magnetoconductivity of two-dimensional Dirac cones and gapped nodal-rings under impurity-potentials in the ultraquantum limit}

\author{Ipsita Mandal}
\email{ipsita.mandal@snu.edu.in}

\author{Sanskar Sharma}

\affiliation{Department of Physics, Shiv Nadar Institution of Eminence (SNIoE), Gautam Buddha Nagar, Uttar Pradesh 201314, India}

\begin{abstract}
We investigate the dc magnetoconductivity of two-dimensional Dirac cones and gapped nodal rings (GNRs) subjected to a perpendicular magnetic field, which quantises the electronic spectrum into Landau levels (LLs). Working in the ultraquantum limit, where only the lowest LL (LLL) is partially occupied, we employ the Kubo--Bastin formalism to compute the transport coefficients for pointlike, Gaussian, and Yukawa impurity-potentials. For the Dirac case, the longitudinal conductivity is field-independent for pointlike impurities and a monotonic function of $B$ for the Gaussian and Yukawa potentials, while the Hall conductivity vanishes identically owing to the particle-hole symmetry of the two neighbouring LLs. The GNR case is qualitatively different: its non-monotonic stretched-checkmark LL spectrum causes the effective LLL to migrate to successively lower indices as the field increases, producing a pronounced oscillatory structure in both conductivities. The longitudinal response develops resonant peaks at LLL degeneracies, while the Hall conductivity traces out a sawtooth pattern with sharp zero-crossings at these same points. These results establish distinct transport fingerprints for the two systems in the extreme quantum limit, and provide a theoretical framework for interpreting magnetotransport experiments on GNRs.
\end{abstract}

\maketitle

\tableofcontents


\section{Introduction}
\label{sec:intro}

Magnetoresistance, the change in a material's resistivity under an applied magnetic field ($\boldsymbol B$), is normally quadratic and small. Departures from this behaviour are a sensitive probe of the underlying electronic structure. Nodal-point semimetals occur in both two and three dimensions \cite{castro_neto_graphene_review, armitage_weyl_dirac_review} --- with graphene as the archetypal two-dimensional (2D) realisation, and Weyl and Dirac semimetals as its three-dimensional (3D) counterparts --- which harbour band-crossing points in their Brillouin zone (BZ). At weak fields, a negative magnetoresistance can arise from the chiral anomaly, a phenomenon specific to the three-dimensional (3D) case and tied to the Berry curvature carried by the nodes \cite{son_spivak_chiral, ips-internode}. Landau quantisation is present at any nonzero field --- for Dirac-like quasiparticles, the spacing between consecutive LLs depends on their energy $E_n$ itself. At low fields, this spacing is small compared to $E_n$, so that quantisation effects can be neglected and a semiclassical Boltzmann treatment suffices. As the field is increased, this spacing becomes resolvable and Shubnikov--de Haas oscillations appear, eventually giving way to the observable quantum-Hall effect \cite{klitzing_qhe}. Returning to the 2D case, which is the focus of this paper, such systems have long been the natural setting for probing this high-field regime: the discovery of 2D magnetotransport in the extreme quantum limit \cite{tsui_stormer_gossard} established that, once the field is strong enough to depopulate all but the lowest Landau level (LLL), the system's transport properties are governed entirely by the physics of that single level. In graphene, whose massless Dirac quasiparticles quantise into LLs with the characteristic $E_n\propto\mathrm{sgn}(n)\sqrt{|n|}$ spacing (rather than the equally spaced ladder of an ordinary 2D electron gas), the distinct character of the extreme quantum limit has been studied extensively \cite{girvin-jach, fuchs-bc-graphene, ando-scba, gusynin06_magneto, goerbig_review, dirac-qtm}. In the ultraquantum limit, reached once only the LLL remains occupied, a positive magnetoresistivity that grows linearly with $B$ emerges, and has long been regarded as a hallmark of 3D Weyl semimetals (WSMs) \cite{abrikosov_qtm_mr}. In Dirac and Weyl systems, the LLL is always associated with the zeroth orbital index ($n=0$), whereas in a typical gapped system, the effective LLL occurs at a field-dependent index $n_g$ \cite{barati-nlsm-qhe}.

Strikingly, a pronounced linear magnetoresistance has now been observed in graphene under ultrahigh fields, with the Fermi level pinned at the zeroth LL, i.e., at the charge-neutrality point \cite{xin_graphene_expt}. Unlike a 3D WSM, where the LLs disperse continuously along the direction of $\boldsymbol B$, graphene's LL spectrum is non-dispersing --- the existing theoretical framework, built for the 3D case on a first-order Born treatment of disorder, was found unable to reproduce the observed behaviour of magnetoresistance, exposing a genuine gap in the theory of Landau-quantised Dirac quasiparticles at strong fields \cite{dirac-qtm}. This gap was closed only recently, by treating the impurity scattering self-consistently \cite{dirac-qtm}. Within a self-consistent Born approximation (SCBA), the disorder self-energy of the LLL and of its two nearest neighbours is considered, and the resulting spectral functions are combined through the Kubo formula to give an analytic longitudinal conductivity in the ultraquantum limit for three archetypal impurity potentials: a $\delta$-function (white-noise) potential, a Gaussian potential, and a Yukawa (screened Coulomb) potential. The $\delta$-function potential reproduces the field-independent minimal conductivity of 2D massless Dirac quasiparticles; the Gaussian potential instead yields a genuinely linear magnetoresistivity whenever its range is smaller than the magnetic length, which Ref.~\cite{dirac-qtm} showed to be in quantitative agreement with the graphene data of Ref.~\cite{xin_graphene_expt}; and the Yukawa potential gives a $B^{-1}$ dependence, qualitatively distinct from the 3D case \cite{dirac-qtm}.

Away from the isotropic untilted Dirac cone (or its generalisations, in two or three dimensions), the associated linear response get modified i the presence of tilting or other anisotropic deformations of the cone (as occurs generically once lattice anisotropy or strain-induced pseudomagnetic fields are included) --- the resulting thermoelectric and magneto-optical coefficients have been studied extensively \cite{ips_tilted_dirac, ips-sanskar, ips-kush, ips-kush-review, yadav23_magneto, ips_cd, ips-ruiz, ips-tilted}. The subset of such studies beyond the weak-$B$ limit, however, remain confined to semimetallic bandstructures whose LL spinors are simple two-component objects at each index $n$: a single, fixed pair of adjacent harmonic-oscillator orbitals, $|n-1\rangle$ and $|n\rangle$, already mixed in the pseudospinorial eigenstate itself, with tilt or anisotropy only reshaping the mixing weights and LL energies. A structurally-richer and equally-natural setting is provided by a gapped nodal-ring (GNR): each LL index $n$ (with $n\geq 0 $) here accompanies two distinct levels, $(n,+)$ and $(n,-)$, rather than the single LL that each $n$ (where $n \in \mathbb{Z}$) labels for a Dirac cone. The 2D versions are proposed to be realised in systems like Kagome-honeycomb lattice \cite{lu_kagome_honeycomb}, Be$_2$C and BeH$_2$ monolayers \cite{yang_lieb_be2c_beh2}, or certain MX (M\,=\,Pd, Pt; X\,=\,S, Se, Te) compounds \cite{jin_mx_nodeline}. In the absence of $\boldsymbol B$, hybridisation between the positive- and negative-energy bands opens a gap $\Delta$ at the ring. The gap persists when a nonzero $\boldsymbol B$ is turned on.

The existing literature on GNRs splits fairly cleanly by field regime and by dimensionality. At low magnetic fields, where LL formation can be neglected and a semiclassical Boltzmann-equation treatment suffices, 3D GNRs, whose Fermi surfaces take toroidal shapes, have been studied for linear response in planar-Hall configurations \cite{yang_review_nlsm, ips-nlsm-ph}. The effects of strain-induced axial gauge fields have also been considered within the same framework \cite{ips-gnr-strain}.
At high fields, by contrast, the LL quantisation by a magnetic field cannot be ignored. In fact, for a 2D GNR, it causes a distinctive non-monotonic ``stretched checkmark'' dependence on the LL index $n$, in clear contrast with the $\sqrt{|n|}$ ladder of a 2D Dirac cone (or the linear-in-$n$ ladder of an ordinary 2D electron gas). Using the Kubo formula in the clean limit, the real and imaginary parts of the longitudinal and Hall optical conductivities have been obtained in closed form for this LL spectrum \cite{barati-nlsm-qhe}. As $B$ is varied at fixed chemical potential ($\upmu$), LLs from one side of the spectrum move below the Fermi level and get filled. LLs from the other side move out and get vacated instead. Consequently, the zero-frequency (dc) Hall conductivity swings back and forth between two consecutive integer Chern numbers $\mathcal C$, giving rise to bumpy quantum-Hall plateaux. The longitudinal optical conductivity likewise displays distinct series of Shubnikov--de Haas peaks, which are associated with interband and intraband transitions among the descending and ascending sets of LLs. This LL problem, however, has been solved only in the clean limit --- dc conductivity arising from disorder-broadening in the ultraquantum-limit has not been addressed. It is precisely this high-field disordered regime on which we focus in the present work. The GNR's LL eigenstates comprise a pair of electron and hole branches built from the \emph{same} orbital wavefunction (indexed by an integer $n \geq 0$), but differing by an extra $s$-dependent index (where $s=\pm$). Hence, it is not obvious a priori how the resulting self-energies organise themselves into the dc conductivity in the ultraquantum limit.

Our aim is to compute the dc conductivity, both longitudinal and Hall, of the 2D Dirac cone and of the GNR in the ultraquantum limit, under impurity-potentials. For this, we use the Kubo--Bastin formalism (described in Sec.~\ref{sec:kubo}), the method of choice for disorder problems of this kind. For the Dirac cone the LLL sits fixed at $n=0$, and Ref.~\cite{dirac-qtm} has already obtained the longitudinal conductivity there for three archetypal impurity-potentials; what remains open, and what we supply here, is the Hall conductivity for the same system. The GNR is a richer problem still, since its LLL is not pinned at $n=0$: its checkmark-shaped spectrum means that the LL index $n_g$ nearest to the chemical potential depends on $B$ itself, through the dimensionless ratio $\rho = 2\,\varepsilon_r/\varepsilon_c$, where $\varepsilon_r$ is the energy scale set by the radius of the parent (ungapped) nodal-ring and $\varepsilon_c$ is the cyclotron energy. For special values of $\rho$, two levels can even become degenerate at $n_g$. Since we evaluate the conductivity strictly within the ultraquantum limit throughout, at each value of $B$, we first identify the LLL index $n_g(B)$, together with any accompanying degeneracy. We then apply the Kubo--Bastin formula to that level and its neighbours. As $B$ is cranked up, $n_g(B)$ decreases through successive integers. It is this migration of the effective LLL that produces the pronounced oscillatory structure in both the longitudinal and the Hall conductivity of the GNR, which we compute explicitly for short-ranged as well as long-ranged impurity-potentials. The resulting behaviour of the GNR is qualitatively very different from the smooth monotonic longitudinal magnetoresistivity of the 2D Dirac case in the ultraquantum limit.

The paper is organised as follows. Sec.~\ref{sec:model-ham} introduces the two model Hamiltonians. It works out their Landau-level spectra and eigenstates in a perpendicular magnetic field. Sec.~\ref{secform} sets up the Kubo--Bastin formalism. It also derives the self-consistent Born treatment of disorder for pointlike, Gaussian, and Yukawa impurity-potentials. Sec.~\ref{sec:selrules-ql} derives the relevant velocity matrix-elements. These fix the selection rules that determine which LLs contribute to the conductivity in the ultraquantum limit for the GNR. Sec.~\ref{secres} presents our numerical results for the longitudinal and Hall conductivities of the GNR, and compares them against the Dirac case. Sec.~\ref{seccon} summarises our findings and outlines some natural extensions. The paper closes with three appendices. Appendix~\ref{appcond} derives the general Kubo--Bastin conductivity formula used throughout the main text. Appendix~\ref{app:sigmaII} relates this formula to the clean-limit Fermi-sea contribution $\sigma_{xy}^{\rm II}$, computed in Ref.~\cite{barati-nlsm-qhe}. Appendix~\ref{appdirac} works out the full conductivity of the Dirac cone, including the vanishing of its Hall response.

We work in natural units, $\hbar=k_B=c = 1$, throughout. We retain the electron charge $e$ explicitly, as a bookkeeping device to track the electric-charge factors.

\section{Model Hamiltonians}
\label{sec:model-ham}

The two systems considered in this work are described by two-band Bloch Hamiltonians in a pseudospin basis. For the isotropic 2D Dirac cone, the Hamiltonian is
\begin{align}
H_{\rm D} (\boldsymbol k) = v_F \, \boldsymbol k \cdot \boldsymbol \sigma
= v_F \, ( k_x \, \sigma_x + k_y \, \sigma_y )\,,
\end{align}
where $v_F$ is the Fermi velocity and $\sigma_{x,y,z}$ are Pauli matrices acting on the sublattice pseudospin \cite{dirac-qtm}. Diagonalising $H_{\rm D}(\boldsymbol k)$ gives the linear spectrum $E(\boldsymbol k) = \pm \, v_F \, |\boldsymbol k|$, touching zero at the single point $\boldsymbol k = 0$, the Dirac node.

The 2D GNR, by contrast, is described by
\begin{align}
H_{\rm GNR} (\boldsymbol k) = \frac{k^2 - k_0^2}{2 \, m^*} \, \sigma_x + \Delta \, \sigma_z \,,
\end{align}
where $m^*$ is the band mass and $\Delta$ is the gap opened at the ring by hybridisation between the two bands \cite{barati-nlsm-qhe}. For $\Delta = 0$, the two bands touch on the ring $k = k_0$, giving the gapless nodal-ring dispersion; a nonzero $\Delta$ gaps this ring uniformly, giving the eigenvalues
\begin{align}
E (\boldsymbol k) = \pm \, \sqrt{ \left( \frac{k^2 - k_0^2}{2 \, m^*} \right)^2 + \Delta^2 } \,.
\end{align}

\subsection{Landau quantisation}
\label{sec:LL-general}

We take the applied field to be $\boldsymbol B = B\,\hat{\boldsymbol z}$, perpendicular to the 2D plane in which either system resides. Minimal coupling is implemented as $\boldsymbol k \to \boldsymbol k + e\,\boldsymbol A$, consistent with the current operator $\boldsymbol j = -e\,\nabla_{\boldsymbol k} H(\boldsymbol k)$ used throughout [cf. Eq.~\eqref{eq:KG}], where $-e$ is the electron charge. We work in the Landau gauge, $\boldsymbol A = B\,(-y,0,0)$, which preserves translational invariance along $x$, so that $k_x$ remains a good quantum number and the eigenstates carry a plane-wave factor $e^{i\,k_x\,x}/\sqrt{L_x}$.

It is convenient to introduce the magnetic length $\ell_B = 1/\sqrt{e\,B}$ (natural units, $\hbar=1$) and the dimensionless guiding-centre coordinate
\begin{align}
\xi = \frac{y}{\ell_B} - \ell_B \, k_x \,.
\end{align}
In terms of these, the minimally-coupled momenta combine into bosonic ladder operators,
\begin{align}
a = \frac{1}{\sqrt 2}\,(\xi + \partial_\xi) \,, \quad
a^\dagger = \frac{1}{\sqrt 2}\,(\xi - \partial_\xi) \,, \quad
[a,a^\dagger] = 1 \,, \quad
a^\dagger a\,\phi_n(y) = n\,\phi_n(y)\,,
\end{align}
where 
\begin{align}
\label{eq:phi-n-def}
\phi_n(y) = \frac{1}{(\pi \, \ell_B^2)^{1/4}} \frac{1}{\sqrt{2^n \,n!}} 
\;H_n\left(\frac{y}{\ell_B}\right) \; \exp\left(-\frac{y^2}{2 \,\ell_B^2}\right).
\end{align}
$\phi_n(y)$ is the harmonic-oscillator wavefunction, written in terms of the Hermite polynomial ($H_n$), and normalised such that $\int dy \, [\phi_n(y)]^2 = 1$. Both $H_{\rm D}(\boldsymbol k)$ and $H_{\rm GNR}(\boldsymbol k)$ depend on $\boldsymbol k$ only through the combinations $k_x + e\,A_x \pm i\,k_y$. These combinations become proportional to $a^\dagger$ and $a$, respectively. Substituting these identifications turns each Bloch Hamiltonian into a $2\times2$ matrix in $(a,a^\dagger)$, acting within a fixed $k_x$ sector. Diagonalising it then yields the LLs.

For the Dirac cone, this substitution gives $H_{\rm D} \to \omega_B\,(a^\dagger\,\sigma_- + a\,\sigma_+)$ with $\sigma_\pm = (\sigma_x \pm i\,\sigma_y)/2$ and $\omega_B = v_F\,\sqrt{2\,e\,B}$, reproducing the LL spectrum $E_{n,s} = s\,E_n$ with
\begin{align}
E_n = \omega_B\,\sqrt{n} \,, \quad n\in 0  \cup \mathbb{Z}^+ \,, \quad s = \pm \,,
\end{align}
and eigenspinors
\begin{align}
\label{eq:dirac-eigenspinor}
|k_x,n,s\rangle = \frac{e^{i\,k_x\,x}}{\sqrt{L_x}}\,
\frac{1}{\sqrt{2}}
\begin{bmatrix} \phi_{n-1}(y) \\ s\,\phi_n(y) \end{bmatrix}  \,.
\end{align}
The anomalous zeroth LL carries no $s$ label (as there is only one eigenspinor at $n=0$), with $E_0=0$ and
\begin{align}
\label{eq:dirac-eigenspinor-LLL}
|k_x,0\rangle = \frac{e^{i\,k_x\,x}}{\sqrt{L_x}}\,(0,\,\phi_0(y))^{\rm T}\,.
\end{align}

For the GNR, the same substitution replaces $k^2 \to \varepsilon_n \equiv \varepsilon_c\,(n+\tfrac12)$, with $\varepsilon_c = e\,B/m^*$ the cyclotron energy, so that $H_{\rm GNR} \to (\varepsilon_n-\varepsilon_r)\,\sigma_x + \Delta\,\sigma_z$ within each $n$-sector, where $\varepsilon_r = k_0^2/(2\,m^*)$. Diagonalising this $2\times2$ matrix gives the spectrum $E_{n,s} = s\,E_n$ with
\begin{align}
\label{eq:gnr-spectrum}
& E_n = \sqrt{ \frac{\varepsilon_c^2} {4}\, ( 2\,n+ 1- \rho )^2+\Delta^2}\,,\quad
\rho = \frac{2\,\varepsilon_r} {\varepsilon_c} \,, \quad n\in 0  \cup \mathbb{Z}^+ \,,
\quad s =\pm\,,
 \quad \varepsilon_r = \frac{k_0^2} {2 \, m^*} \,, \quad \varepsilon_c =  \frac{e \, B} {m^*}\,,
\end{align}
and eigenspinors
\begin{align}
\label{eq:gnr-eigenspinor}
|k_x,n,s\rangle = \frac{e^{i\,k_x\,x}}{\sqrt{L_x}} \,\frac{1}{\sqrt 2}
\begin{bmatrix} 
A_n^{s+} \\ s\,\mathrm{sgn}(\sin\theta_n)\,A_n^{s-} \end{bmatrix} \phi_n(y) \,,
\quad A_n^{s\pm} = \sqrt{1\pm s\cos\theta_n } \,.
\end{align}
Here, $\theta_n = \arctan[(\varepsilon_n-\varepsilon_r)/\Delta]$ is the mixing angle, satisfying $\cos\theta_n = \Delta/E_n$ and $\sin\theta_n = (\varepsilon_n-\varepsilon_r)/E_n$.

The two spectra differ qualitatively because $\varepsilon_n \propto n+\tfrac12$ enters $E_n^{\rm GNR}$ quadratically through $(\varepsilon_n-\varepsilon_r)^2$, whereas $\omega_B\sqrt{n}$ enters $E_n^{\rm D}$ directly: the Dirac ladder is monotonic in $n$, while the GNR ladder is non-monotonic in $n$, first decreasing and then increasing as $\varepsilon_n$ sweeps past $\varepsilon_r$. The latter resembles a ``stretched checkmark'' pattern discussed in Ref. \cite{barati-nlsm-qhe}.

\subsection{Unique non-monotonic structure of GNR's LLs}
\label{sec:selrules}

Fig.~\ref{figll} represents the behaviour of the GNR's LLs as functions of $\rho$.
 Although the LL spectrum of GNR is symmetric about zero of energy, satisfying $E_{n,s=+} = -E_{n,s=-}$, there is no LL at the zero value. The LL with the lowest positive value (abbreviated as LLL) appears at the minimum of $ f \equiv | \rho - 2\,n-1 |$, given that $n$ can only be zero or a positive integer. Therefore, it resides at index $n_{g}= \mathrm{nint}( \tfrac{\rho -1}{2})$ (also denoted here as $ n_g \equiv n_{\rm LLL}$), where $ \mathrm{nint}(x)$ is the integer closest to $x$ (half-integers are rounded down). 
Depending on $\rho$, the LLs may be non-degenerate or doubly-degenerate:
\begin{enumerate}

\item If $ \rho$ is non-integer, there is no degeneracy of any LL.

\item If $ \rho =  2 \, N +1 $ (odd integer) with $N \in 0 \cup \mathbb{Z}^+$, then $\rm{min} (f)=0$, $n_g = N$, and $ E_g =  \Delta $. For this case, $ ( 2\,n+ 1- \rho )^2 /4 = (n-N)^2$, leading to $(N+N'-N)^2 = (N-N'-N)^2 $ --- this implies $ E_{n_g + N'} = E_{n_g - N'}$. Therefore, as long as $n_g - N'\geq 0 $, we have doubly-degenerate LLs flanking the two sides of the LLL.
One can easily observe this characteristic in Fig.~\ref{figll}. For $n_g - N'< 0 $ (or $N' > n_g$), $ E_{n_g + N'}$ is not degenerate with any other LL. 

\begin{figure}[t!]
    \centering
\includegraphics[width= 0.75\linewidth]{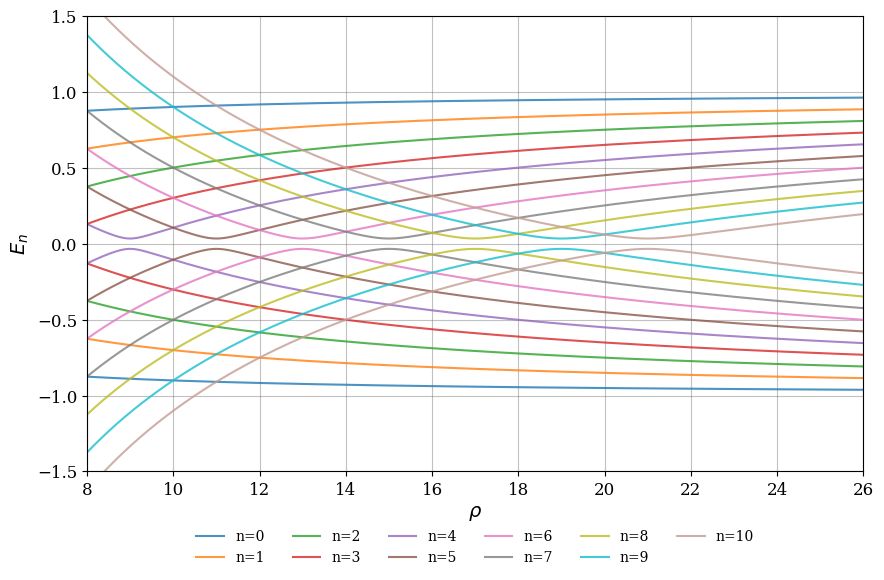}
\caption{LLs of GNR as functions of $\rho$, setting $m^* = 2 \times 10^4 $ eV, $k_0 = 200 $ eV, and $\Delta = 0.034 $ eV.\label{figll}}
\end{figure}

\item If $ \rho = 2\, N$ (even integer) with $N \in 0 \cup \mathbb{Z}^+$, then $n= N-1$ and $n = N $ give same energy because $[2\,(N-1)+1 -\rho ]^2  = [2\,N+1 -\rho ]^2 = 1 $. This gives a doubly-degenerate LLL where 2 LLs overlap with indices $n_1 = N-1 $ and $n_2  =  N = n_1+1 $. For this case, $ ( 2\,n+ 1- \rho )^2 = [ 2\,(n-N)+1 ]^2$, leading to $ [ 2\,(N+N'-1-N)+1 ]^2 = [ 2\,(N-N'-N)+1 ]^2 = (2\, N' -1)^2$ --- this implies $ E_{n_1 + N'} = E_{n_2 - N'}$ or $ E_{n_1 + N'} = E_{n_1+1 - N'}$.
Therefore, as long as $n_1+1 - N' \geq 0 $, we have doubly-degenerate LLs flanking the two sides of the doubly-degenerate LLLs.
One can easily observe this feature in Fig.~\ref{figll}. For $ n_1+1 - N'< 0 $ (or $N' > n_1 + 1$), $ E_{n_1 + N'}$ is not degenerate with any other LL.

\end{enumerate}
Accounting for possible degeneracy, we denote the LLL-indices by the set $\lbrace n_g \rbrace$ with $s=+$, each with energy $E_g$ (the lowest non-negative value of LL-energy). The set of LLs closest in energy to $E_g$ is indicated by the sets $\lbrace n_{g_n} \rbrace$, with either $s=+$ or $s=-$ possible.

\subsection{Contributions in the ultraquantum limit}
\label{sec-comp}

In Dirac systems, the LLL has two neighbouring levels, $(n=1, s =\pm )$, which are equidistant from $E_0=0$. Thus, both $ \pm E_{1}$contribute to $\sigma_{\mu \nu}$ [defined in Eq.~\eqref{eq:KB}] equally in the ultraquantum limit for the Dirac case. On the contrary,
the GNR exhibits a fundamentally different LL structure --- the neighbours (e.g., $n_g \pm 1$) are not in general equidistant from $E_g$ because the LL spacing is asymmetric about the LLL. As we have discussed, the LL energies form a non-monotonic stretched-checkmark pattern that directly determines which neighbours participate in transport. At weak magnetic fields, the LLL corresponds to a high-index LL. As the field increases, the LLL successively shifts to lower-index LLs, with each LL becoming the LLL over a finite field range before passing this role to the next lower-index LL.

Thus, the selection of which neighbours dominate the conductivity must be determined by direct comparison of energy spacings $\Delta_+ = E_{n_g+1} - E_g$,  $\Delta_- = E_{n_g-1} - E_g$, and $E_{n_g} - (-E_g)$.  This asymmetry is the defining feature of GNR transport in the quantum limit. The conductivity formula retains the Bastin structure from Sec.~\ref{sec:kubo}, but the identity of the participating levels depends on the spectrum parameters $\rho$, $\varepsilon_r$, $\Delta$. Sec.~\ref{sec:selrules} classifies these regimes.

\section{Formalism to compute conductivity}
\label{secform}

The conductivity tensor for a generic quantum system, in the linear-response regime, can be evaluated using the Kubo formalism~\cite{kubo1957, bruus}:
\begin{align}
\sigma_{\mu\nu}(\omega) = \frac{1}{\mathcal V }
\int_0^\infty dt\,e^{-i\omega t} \int_0^\beta d\lambda\,\left\langle
j_\nu(-i\,\lambda)\,j_\mu(t) \right\rangle ,
\end{align}
where $\mathcal V$ is the system volume, $\mu,\nu\in\{x,y,z\}$ denote the Cartesian directions, $\omega$ is the probe frequency, $\beta=1/T$ is the inverse temperature, and $\mathcal H$ is the system Hamiltonian. The current operator evolves as
\begin{align}
j_\mu(t) =e^{i \, \mathcal H \, t} \, j_\mu \, e^{-i\,\mathcal H\, t} \,, \quad
j_\nu(-i\,\lambda) =e^{\lambda \,\mathcal H} \,j_\nu \,e^{-\lambda \,\mathcal H}\,,
\end{align}
and $\langle\cdots\rangle=\mathrm{Tr}(e^{-\beta\mathcal H}\cdots)/Z$ denotes the thermal equilibrium average, with $Z=\mathrm{Tr}(e^{-\beta \,\mathcal H})$. Throughout this work, we use natural the units with $\hbar = k_B=1$.
We note that while $j_\mu(t)$ is the Heisenberg-picture current operator, $j_\nu(-i \,\lambda) $ is its imaginary-time counterpart. The angular brackets denote the thermal equilibrium expectation value with respect to $ \mathcal H$.
For a translationally invariant system, $\mathcal H \rightarrow H(\boldsymbol{k})$ denotes the Bloch Hamiltonian in the momentum space.
In the non-interacting limit, the current operator in the Heisenberg picture can be expanded in the basis of single-particle eigenstates $\{|n\rangle\}$ of $\mathcal{H}$ as
\begin{align}
j_\mu(t) = \sum_{n,m}\langle n|j_\mu|m\rangle\, c_n^\dagger\, c_m\,e^{i\,(E_n-E_m)t}\,,
\end{align}
where $E_n$ denotes the energy of the eigenstate $|n\rangle$, and $c_n^\dagger$ ($c_n$) is the fermionic creation (annihilation) operator for eigenstate $|n\rangle$. Substituting this expression into the Kubo formula and evaluating the thermal average yields the Kubo--Greenwood formula \cite{greenwood},
\begin{align}
\sigma_{\mu \nu}(\omega)
=\frac{i}{\mathcal V}\sum_{n,m}
\frac{f_n-f_m}{E_m-E_n} \; \frac{ \langle n|\,j_\nu\,|m\rangle \,\langle m| \, j_\mu\, |n\rangle }
{\omega+i \, \eta-( E_m- E_n)}\,.
\end{align}
Here, $f(E) = [e^{\beta(E-\mu)}+1]^{-1}$ is the Fermi--Dirac distribution, $f_n \equiv f (E_n)$, $\eta\rightarrow0^+$ is a positive infinitesimal quantity, and $\boldsymbol{j} = -e\,\nabla_{\boldsymbol k}   H(\boldsymbol k)$ [for a Bloch Hamiltonian $ H(\boldsymbol k)$]. Thus, the above formula expresses the conductivity as a sum over pairs of exact eigenstates $|m\rangle$, $|n\rangle$ of the full Hamiltonian, weighted by their Fermi--Dirac occupancies \cite{mermin}.

Taking the real part by using $\frac{1}{u +i\,\eta}= \mathcal{P}\!\left(\frac{1}{u}\right)-i\,\pi\,\delta(u)$, we get
\begin{align}
\mathrm{Re}[\sigma_{\mu \nu}(\omega)]
=\frac{\pi \, e^2}{\mathcal V} \sum_{m,n} \frac{f_m-f_n}{E_n-E_m}\,
\langle n|\,v_\nu\,|m\rangle \,\langle m| \, v_\mu\, |n\rangle
\, \delta\!\left(\omega - (E_n-E_m)\right),
\label{eq:KG}
\end{align}
where $v_\mu $ is the velocity operator associated with $H(\boldsymbol k)$ (equivalently, $j_\mu = - \, e\,v_\mu $). Once we specialise to the dc limit, only the real (dissipative) part contributes and, henceforth, we simply write  $\sigma_{\mu \nu}$ for $\mathrm{Re}[\sigma_{\mu \nu}] $. Eq.~\eqref{eq:KG} is the Kubo--Greenwood representation of the conductivity tensor and is formally exact for both clean and disordered systems. In a disordered system, however, the states $|m\rangle$ and $|n\rangle$ are exact eigenstates of the realisation-dependent Hamiltonian $\mathcal H + V_{\mathrm{disorder}}$ and are generally inaccessible analytically. Moreover, physical observables require averaging over
disorder realisations. It is therefore advantageous to reformulate the
conductivity in terms of disorder-averaged Green's functions. This leads
to the Kubo--Bastin formalism \cite{bastin1971} discussed below.

\subsection{Kubo--Bastin description}
\label{sec:kubo}

The disorder-averaged Green's functions are evaluated within the self-consistent Born approximation (SCBA)~\cite{ando-scba, giamarchi-scba}, which is appropriate for weak short-range impurity scattering. Within SCBA we retain only the bare-bubble contribution to the current-current correlation function, while neglecting ladder vertex corrections.\footnote{Ladder diagrams describe repeated impurity scattering events that renormalise the current vertex; summing them is equivalent to solving the Bethe--Salpeter equation for the dressed current vertex~\cite{mahan, bruus}.} For weak disorder in a magnetic field and at $T=0$ (or more generally when $T/\upmu\ll 1$, where a Sommerfeld expansion is applicable), these corrections are subleading~\cite{mermin}. Consequently, the dressed current vertex is approximated by the bare current operator, $\boldsymbol{j}$. Within these approximations, the dc conductivity assumes the Kubo--Bastin form, which follows from the standard Kubo formula \cite{kubo1957} expressed in terms of disorder-averaged Green's functions \cite{bruus, mahan, rammer},
\begin{align}
\label{eq:KB}
\sigma_{\mu \nu} = \frac{e^2}{2\,\pi\, \mathcal{V}} 
\int dE \left( -\frac{\partial f}{\partial E} \right)
\mathrm{Tr} \!\left[ v_\nu \,G^R(E) \,v_\mu \,G^A(E) \right] .
\end{align}
Here, $\sigma_{\mu\nu}$ is again understood as $\mathrm{Re}[\sigma_{\mu\nu}]$, per the dc/dissipative-part convention established after Eq.~\eqref{eq:KG}; the trace itself is generally complex, and only its real part is retained (see Appendix~\ref{appcond}). $G^{R}$ ($G^{A}$) denotes the disorder-averaged retarded (advanced) Green's function. 

To make contact with the LL description, we introduce the spectral function,
\begin{align}
\mathcal A_n(E) \equiv -2\,\mathrm{Im} [G_n^R(E) ] = i\,\bigl[G_n^R(E) - G_n^A(E)\bigr]\,,
\end{align}
whose width is set by the disorder-induced broadening of the $n$th LL. The width of $A_n$ is characterised by the disorder broadening $ \Gamma \sim 1/\tau $ (where $\tau$ is the scattering lifetime).
Within the LL-overlap approximation at $T=0$, Eq.~\eqref{eq:KB} leads to (see Appendix~\ref{appcond} for more details)
\begin{align}
\label{eq:sig-general}
& \sigma_{xx} (\upmu) = \frac{e^2}{8\,\pi\,\mathcal V}
\sum_{n,m}|\langle n|v_x|m\rangle|^2\Big[\mathcal A_m(\upmu)\,\mathcal A_n(\upmu) + \mathcal B_m(\upmu)\,\mathcal B_n(\upmu)\Big]\text{ and }\nn
& \sigma_{xy} (\upmu)  = -\,\frac{e^2}{8\,\pi\,\mathcal V}\sum_{n,m}\,\mathrm{sgn}(m-n)\,
|\langle n|v_x|m\rangle|^2\,
\Big[\mathcal B_m(\upmu)\,\mathcal A_n(\upmu) - \mathcal A_m(\upmu)\,\mathcal B_n(\upmu)\Big]\,,
\end{align}
where
\begin{align}
\mathcal{A}_{n} ( \upmu) = \frac{2 \,\Gamma_{n}(\upmu)} { D_n^2(\upmu)+\Gamma_{n}^2(\upmu) }\,,\,,\quad
\mathcal B_{n}(\upmu) = \frac{2\,D_n(\upmu)}{D_n^2 (\upmu) +\Gamma^2_n(\upmu)}\,,
\quad
D_n(\upmu) \equiv \upmu - E_n - \mathrm{Re}[\Sigma^R_n(\upmu)]\,,\quad
\Gamma_{n}(\upmu) =  - \,\mathrm{Im}[\Sigma^R_{n}(\upmu) ] \,,
\end{align}
and Eq.~\eqref{eq:vyvx_matrix_element} (derived later, in Sec.~\ref{sec:selrules-ql}, for the GNR model) has been used. The form of $\sigma_{xy}$ is not specific to the gapped nodal-ring model. It follows from the Kubo--Bastin formula once the velocity matrix-elements satisfy Eq.~\eqref{eq:vyvx_matrix_element}. An analogous relation is obtained for Dirac LLs --- so the same algebraic structure of $\sigma_{xy}$ applies there as well, with model dependence entering only through the LL spectrum and matrix-elements.
Here, $\Sigma^R_{n}(\upmu)$ is the retarded self-energy of the $n$th LL. Henceforth, we will drop the subscript $R$ from $\Sigma^R_{n}$ to avoid cluttering of notations.

In the ultraquantum limit, where the lowest Landau level (LLL) is partially occupied and $ \upmu \simeq E_g$ (where $E_g$ is the energy of the LLL), the effective integration window is set by the disorder-induced broadening $\Gamma$ rather than thermal smearing: even at finite temperature, if $T \lesssim \Gamma$, thermal effects are subleading and the above forms remain accurate.
Eq.~\eqref{eq:sig-general} is applicable for disorder models that, after ensemble averaging, preserve translational invariance and rotational isotropy, so that the spectral functions remain diagonal in the eigenbasis of the clean Hamiltonian.

\subsection{Disorder-induced self-energy}
\label{secdis}

In order to obtain a physically meaningful quantity, the disorder configurations need to be averaged over an ensemble of disorder realisations. The disorder-averaged Green's function $\langle G_{\alpha\alpha'}(\varepsilon)\rangle$ is related to the unperturbed Green's function $G^{(0)}_\alpha(\varepsilon) = (\varepsilon - \varepsilon_\alpha)^{-1}$ via Dyson's equation \cite{ando-scba},
\begin{align}
\langle G_{\alpha\alpha'}(\varepsilon)\rangle = \delta_{\alpha\alpha'} G^{(0)}_\alpha(\varepsilon)
+ G^{(0)}_\alpha(\varepsilon)\sum_{\alpha''}\Sigma_{\alpha\alpha''}(\varepsilon)\langle G_{\alpha''\alpha'}(\varepsilon)\rangle\,,
\end{align}
where $\alpha \equiv (n, k_x)$ collects all quantum numbers. On disorder-averaging, we get the so-called impurity correlator,
\begin{align}
\overline{ V (\boldsymbol{r}) \,V (\boldsymbol{\tilde r} ) } = 
n_{\rm imp}  \int d^2\boldsymbol{q} \,
e^{i\, \boldsymbol{q}\cdot (\boldsymbol{r} - \boldsymbol{\tilde r})}
u (\boldsymbol{q} ) \,u (- \boldsymbol{q} )\,.
\end{align}
We consider scatterers inducing an isotropic potential for the electrons, such that $u (\boldsymbol{q} ) =  u ( |\boldsymbol{q}|)$.
We define the matrix-element,
\begin{align}
\label{eqmat00}
\mathcal{M}_{k_x, k_x'}^{n, n'} = \overline{\langle k_x', n'| V(\boldsymbol{r}) \, |k_x, n \rangle \, 
 \langle k_x, n| V (\boldsymbol{\tilde r}) \, |k_x', n' \rangle } \,,
\end{align}
which appears while computing the self-energy.
Within the self-consistent Born approximation (SCBA), the proper self-energy is \cite{ando-scba}
\begin{align}
\label{eqselinit}
\Sigma_{n} (\varepsilon) & = L_x \sum_{n'} \int \frac{ dk_x'} {2 \, \pi}  \;
 \mathcal{M}_{k_x, k_x'}^{n, n'}  \; G_{n'} (\varepsilon) \,,
\text{ where }
 G_n (\varepsilon) = \frac{1}{ \varepsilon -  \varepsilon_{n}- \Sigma_{n} (\varepsilon)} \,.
 \end{align}
A crucial observation for our impurity-models is that, because translational invariance is recovered after the impurity average, $\Sigma$ becomes diagonal in $k_x$ and furthermore becomes diagonal in the LL index $n$ \cite{ando-scba,giamarchi-scba}. The integral feeding into $\Sigma_{n}(\varepsilon)$ then samples only the dispersion, leaving $\Sigma_n$ and $G_n$ as functions of $\varepsilon$ alone. In addition, vertex corrections to the conductivity vanish identically for short-range (white-noise) scatterers \cite{ando-scba}, so the Kubo formula reduces directly to the spectral form of Sec.~\ref{sec:kubo} with no ladder-diagram resummation required.

In our calculations in the ultraquantum limit, we assume that a positive chemical potential ($\upmu > 0 $) is applied such that it is close to the bottom of the LLL with energy $E_g$ (i.e., $ \upmu \simeq E_g$). Then the main contribution of the sum over $n'$ in Eq.~\eqref{eqselinit} comes from the term $n' = n_g$. We thus single out the terms with $n = n_g$ from the sum and evaluate the self energy if the LLL by considering the dominant term of the summation, evaluated at the single energy $\upmu$ rather than as an energy-integrated or thermally smeared quantity [this is justified by the $T\to0$ collapse of the energy integral onto $E=\upmu$; see Appendix~\ref{appcond}]:
\begin{align}
\label{eqsel_LLL}
\Sigma_{n_g}(\upmu) \approx
 L_x \int \frac{ dk_x'} {2 \, \pi}  \;
 \mathcal{M}_{k_x, k_x'}^{n_g, n_g}  \; G_{n_g} (\upmu) 
 = L_x \int \frac{ dk_x'} {2 \, \pi}  \;
\frac{ \mathcal{M}_{k_x, k_x'}^{n_{g}, n_g } }
{\upmu -  E_g - \Sigma_{ n_g} (\upmu)}
\approx L_x \int \frac{ dk_x'} {2 \, \pi}  \;
\frac{ \mathcal{M}_{k_x, k_x'}^{n_{g}, n_g } }
{- \Sigma_{ n_g} (\upmu)}  \,.
\end{align}
From this equation, we solve for $\Sigma_{n_g}(\upmu)$, which appears both on the left-hand side (LHS) and right-hand side. Next, while considering the contributing terms to the conductivity, we consider the terms involving the LLL itself and those involving to a higher-energy level closest to the LLL (say, $n= n_{gn}$). We will approximate its self-energy by the dominant term at $\upmu \simeq E_g $, obtained by retaining only the $n'= n_g$ term of the summation in Eq.~\eqref{eqselinit}:
\begin{align}
\label{eqsel_neigh}
\Sigma_{n_{gn}}(\upmu) \approx
 L_x \int \frac{ dk_x'} {2 \, \pi}  \;
 \mathcal{M}_{k_x, k_x'}^{n_{gn}, n_g}  \; G_{n_g} (\upmu) 
 = L_x \int \frac{ dk_x'} {2 \, \pi}  \;
\frac{ \mathcal{M}_{k_x, k_x'}^{n_{gn}, n_g } }
{\upmu -  E_g - \Sigma_{ n_g} (\upmu)}
 \approx L_x \int \frac{ dk_x'} {2 \, \pi}  \;
\frac{ \mathcal{M}_{k_x, k_x'}^{n_{gn}, n_g } }
{ - \Sigma_{ n_g} (\upmu)}\,.
\end{align}
It is justified as long as we are assuming well-separated LLs and $\upmu$ sufficiently close to $E_g$, so that the dominant spectral weight coming from $ G_{n_g}(\upmu)$ dominates the summation. Physics-wise, the approximated formula is a statement that the linewidth of the neighbouring LL is generated predominantly by scattering into the LLL, whose spectral weight is largest near $\upmu $. 

Here, some comments are in order: In our prescription, the dominant term is picked up for computing the sum in the self-energy shown in Eq.~\eqref{eqselinit}. This led to an SCBA form for $n_g$ LLL, but not for $n_{gn}$, as $\upmu$ cuts $E_g$ (and not $E_{ng}$). This is important to note --- $n_{gn}$ contributes due to spectral broadening (spilling into the broadened spectrum of $n_g$) and $\Sigma_{n_{gn}}(\upmu)$ is not to be solved self-consistently. This point was missed in the computations of Ref.~\cite{dirac-qtm} and they ended up with the incorrect value of $\Sigma_{n_{gn}}$. We have corrected it in this paper.

For our two-band Hamiltonians, the quantum numbers must be extended to $\alpha \equiv (n, s, k_x)$ with $s = \pm$ labelling the two branches $\pm E_n$ of the nodal-ring spectrum. The SCBA self-energy at this level of generality reads
\begin{align}
\Sigma_{n,s;\,n',s'}(\varepsilon)& = L_x \sum_{n'',s''} \int \frac{dk_x'}{2 \, \pi}\;
\mathcal{M}^{n,s;\,n'',s''}_{k_x,k_x'}\; G_{n'',s''}(\varepsilon)\,,\quad
G_{n,s}(\varepsilon) = \frac{1}{\varepsilon - s\,E_n - \Sigma_{n,s;\,n,s}(\varepsilon)}\,,
\end{align}
and $\mathcal{M}^{n,s;\,n'',s''}_{k_x,k_x'}$ is the generalisation of Eq.~\eqref{eqmat00} to include the $s$ index [cf.\ Eq.~\eqref{eqmat}]. After disorder averaging, translational invariance forces $\Sigma_{n,s;\,n',s'}$ to be diagonal in $k_x$. It is further diagonal in both $n$ and $s$, recovering Eq.~\eqref{eqselfen}.


\subsubsection{Short-ranged versus long-ranged impurities}
\label{sec:disorder-comparison}

The conductivity formula Eq.~\eqref{eq:KB} applies universally to both short-ranged and long-ranged disorder. However, the physical consequences of disorder type enter through the matrix-elements that feed into the disorder-averaged Green's functions. The selection of which LLs participate in transport (discussed in Sec.~\ref{sec:selrules}) is determined by energy spacing $\Delta_\pm$ and is thus completely insensitive to disorder type. By contrast, the spectral overlap magnitudes and the momentum-transfer structure of scattering differ markedly between short- and long-ranged regimes.

Since the conductivity is evaluated within the bare-bubble approximation, ladder vertex corrections are neglected, and thus intraband vertex renormalisation due to disorder geometry is beyond the scope of the present treatment. Our comparison between short- and long-ranged disorder therefore reflects the interplay of disorder broadening through the Green's functions and the momentum dependence of the scattering potential. The latter is encoded in the impurity matrix-elements $\mathcal{M}_{k_x, k_x'}^{n,n'}$, which depend on the Fourier transform of the disorder potential. For white-noise (short-range) scatterers, this transform is constant; for Gaussian (long-range) scatterers, it falls off exponentially with momentum transfer. We would like to emphasise that the disorder self-energy remains diagonal in the band index for the GNR case, because a scalar disorder does not generate any mixing between the $s = +$ and $ s = -$ sectors (unlike graphene, where intervalley scattering couples the $K$ and $K'$ points). Vertex corrections, however, may still renormalise the current within each sector separately. Since the present work neglects ladder corrections altogether, such intraband vertex renormalisation is beyond the scope of the present treatment.

The physical consequences of the $ R/\ell_B$ ratio are as follows [where $R$ appears in Eq.~\eqref{eq:gaussian-pot}]. When $R\ll\ell_B$, the Gaussian form factor is approximately unity over the characteristic momentum transfers connecting Landau levels, so the disorder effectively behaves as white noise and the short-ranged limit is recovered. As $d$ becomes comparable to $\ell_B$, momentum transfers satisfying $|\boldsymbol{q}|\gtrsim 1/E$ become progressively suppressed, producing a crossover in which forward scattering is enhanced whilst large-angle scattering continues to contribute. In the opposite limit $R \gg\ell_B$, the impurity correlator strongly suppresses large momentum transfers, so that scattering is dominated by small-angle processes. The conductivity magnitude and its dependence on disorder parameters (density, strength, range) are thus sensitive functions of this crossover, even though the LL-selection mechanism remains disorder-independent.

\subsubsection{Pointlike impurities}
\label{sec:sr-imp}

Let us take the simplest case of pointlike impurities, such that the total perturbing potential is created by
\begin{align}
V (\boldsymbol{r}-\boldsymbol{r}_i)= V_0 \,
\delta^{2}(\boldsymbol{r}-\boldsymbol{r}_i)\,,
\end{align} 
leading to the Fourier-transform $u (\boldsymbol{q}) = V_0$.
This model for short-ranged scatterers is referred to as white-noise disorder. The terminology originates from the fact that the impurity-correlator in the momentum space is constant, being equal to $V_0^2$.

Assuming a single-band problem,
\begin{align}
\mathcal{M}_{k_x, k_x'}^{n, n'}  &  
= \frac{ 1 } {L_x^2}\int d \boldsymbol{r} \, d \boldsymbol{\tilde r} \,
 \overline{  V(\boldsymbol{r})\,  V(\boldsymbol{r})' }
 \,e^{i \, (k_x-k_x') \, x} \,  e^{- i \,(k_x - k_x') \, \tilde x}
 \, \phi_{n'} (y-y^0_{k_x'}) \, \phi_{ n} ( y- y^0_{k_x}) \, \phi_{n} (\tilde y-y^0_{k_x})\, 
 \phi_{n'} (\tilde y - y^0_{k_x'})
\nn & = \frac{ n_{\rm imp} \, V_0^2} {L_x^2} \int dx \, dy 
\left[ \phi_{ n'} ( y- y^0_{k_x'}) \right]^2  \left[\phi_{ n} (y- y^0_{k_x} )  \right]^2
= \frac{ n_{\rm imp} \, V_0^2} {L_x} \int  dy
\left[ \phi_{ n'} ( y- y^0_{k_x'}) \right]^2  \left[\phi_{ n} ( y- y^0_{k_x})   \right]^2,
\end{align}
where $ y^0_{k_x'} =  \ell^2_B  \, k_x'$ and $ y^0_{k_x} =  \ell^2_B  \, k_x$.
Using the change of variable $y'=\ell_B^2\,k_x'$ and $\int dy\,\phi_{n'}(y)^2=1$, we get
\begin{align}
\label{eqMpointlike}
\int\frac{dk_x'}{2 \, \pi}\int\frac{dq_y}{2 \, \pi}\,\mathcal{M}^{n,n'}_{k_x,k_x'}
&= \frac{n_{\rm imp}\,V_0^2} {2 \,\pi\, \ell_B^2}\int dy \, [\phi_n(y-y_{k_x}^0)]^2
\int dy' \, [\phi_{n'}(y-y')]^2 = \frac{n_{\rm imp}\,V_0^2} {2 \,\pi \,\ell_B^2}\,,
\end{align}
independent of $n$ and $n'$.

We can evaluate the self-energy as
\begin{align}
\label{eqselfen}
\Sigma_{n} (\varepsilon) & =  n_{\rm imp} \, V_0^2
 \sum_{n'} \int \frac{ dk_x' \, dy } {2 \, \pi}  \;
\frac{ \left[ \phi_{ n'} ( y- \ell^2_B  \, k_x') \right]^2  
\left[\phi_{ n} ( y- \ell^2_B  \, k_x)   \right]^2 }
{ \varepsilon -  \varepsilon_{n'}- \Sigma_{ n'} (\varepsilon)} 
=  \frac{ n_{\rm imp} \, V_0^2} { 2 \,\pi \,\ell_B^2} \sum_{n'} G_{n'} (\varepsilon)\,.
 \end{align}
In the regime of well-separated LLs, if $\varepsilon$ is close to the $N^{\rm th}$ LL, then the leading-order term in the summation on the right-hand side of Eq.~\eqref{eqselfen} is contributed from $n' =  N $ [$i.e., G_{N} (\varepsilon)$]. We can then approximate the self-energy expression as \cite{ando-scba,disordered-Dirac-visco}
\begin{align}
\label{eqselfen1}
\Sigma_N (\varepsilon)
\simeq  \frac{ n_{\rm imp} \, V_0^2} { 2 \,\pi \,\ell_B^2} \,
\frac{1} { \varepsilon -  \varepsilon_N- \Sigma_{ N} (\varepsilon)} \,.
 \end{align}

\subsubsection{Long-ranged impurity potentials: Gaussian}
\label{sec:gauss}

We now turn to the case where the impurity scattering potential extends over length scales that are comparable to or larger than the magnetic length. This situation arises naturally in systems where charged impurities are screened by a finite Debye length, or where the disorder originates from smooth potential variations in the material. We model such long-ranged scattering through a Gaussian potential of the form
\begin{align}
V(\boldsymbol{r} - \boldsymbol{r}_i ) = V_0 \left(\frac{1}{\sqrt{2 \, \pi}\,R}\right)^2 
\, e^{-\frac{ (\boldsymbol{r}-\boldsymbol{r}_i)^2}{2\, R^2}}\,,
\label{eq:gaussian-pot}
\end{align}
where $R$ is the characteristic range and $V_0$ the strength. In the short-range limit of $ R \to 0$, this reduces to the pointlike scatterer $V_0 \, \delta(\boldsymbol{r})$ considered in Sec.~\ref{sec:sr-imp}. 
The Fourier transform is
\begin{align}
u(\boldsymbol{q}) = \int d^2\boldsymbol{r}\, e^{i\, \boldsymbol{q} \cdot \boldsymbol{r}}\, V(\boldsymbol{r}) 
 =  V_0 \, e^{-\frac{\boldsymbol{q}^2 \, R^2} {2} }\,,
\end{align}
which is itself Gaussian in momentum space, sharply-peaked at $\boldsymbol{q} = \boldsymbol{0}$ with width $\sim 1/ R$. This momentum-space structure is the key distinction from the short-range case, where $u(\boldsymbol{q}) = V_0$ is constant: processes involving large momentum transfer $|\boldsymbol{q}| \gg 1/R$ are now exponentially suppressed.

The disorder-averaged two-impurity correlator follows from substituting the Fourier potential into the general expression. Since $u(-\boldsymbol{q}) = u(\boldsymbol{q})$ for a real even potential, we have $u(\boldsymbol{q})\,u(-\boldsymbol{q}) = V_0^2 \,e^{-\boldsymbol{q}^2 \,R^2}$. Writing $\boldsymbol{q} = (q_x, q_y)$, with $q_x = k_x - k_x'$ denoting the momentum transfer in the $x$-direction.
The disorder-averaged two-impurity correlator becomes
\begin{align}
\overline{V(\boldsymbol{r})\,V(\boldsymbol{\tilde r})} = 
n_{\rm imp} \int d^2\boldsymbol{q} \, e^{i\boldsymbol{q} \cdot (\boldsymbol{r} - \boldsymbol{\tilde r})}
\, u(\boldsymbol{q})\,u(-\boldsymbol{q}) = 
n_{\rm imp}\,V_0^2 \int d^2\boldsymbol{q} \, e^{i\,\boldsymbol{q} \cdot (\boldsymbol{r} - \boldsymbol{\tilde r})} 
\, e^{-\boldsymbol{q}^2 \,R^2}.
\end{align}
Because $u(\boldsymbol{q})$ is now $|\boldsymbol{q}|$-dependent, the product $u(\boldsymbol{q})\,u(-\boldsymbol{q}) = V_0^2 \, e^{-\boldsymbol{q}^2 \, R^2}$ is no longer constant as it is for short-range impurities. This $\boldsymbol{q}$-dependence carries forward into the scattering matrix-elements and fundamentally alters how momentum transfers enter into the calculations.

For the Gaussian potential, the Fourier transform introduces an exponential factor in momentum transfer. Following the same procedure as in Sec.~\ref{sec:sr-imp}, the $x$-direction integral performs the momentum-space integration, and the momentum-conserving delta function emerges from the Fourier transform of the spatial coordinates. The key observation is that we separate the $x$ and $y$ components of the momentum transfer: the $x$-component momentum transfer $k_x - k_x'$ couples to the current direction and appears explicitly in the scattering vertex, whilst the $y$-component $q_y$ enters through the spatial overlap of Landau wavefunctions localised at different guiding centres. The result is that the matrix-element $\mathcal{M}_{k_x, k_x'}^{n,n'}$ factorises as
\begin{align}
\mathcal{M}_{k_x, k_x'}^{n,n'} =n_{\rm imp}\,V_0^2\,e^{-\boldsymbol{q}^2 \,R^2}
\left|I^{n,n'}(k_x,k_x')\right|^2 ,\quad
I^{n, n'} (k_x, k_x') = \int_{-\infty}^{\infty} dy \, e^{i \,q_y\, y} \, 
\phi_{n'}(y - y_{k'_x}^0) \, \phi_n(y - y_{k_x}^0)\,.
\label{eq:matrix-elem-factored}
\end{align}
$I^{n,n'}(k_x, k_x')$ represents the overlap of Landau wavefunctions in the $y$-direction, including the $q_y$-dependent parts of both the wavefunction products and the disorder potential. The Gaussian form-factor weights scattering processes inversely with the square of the momentum transfer: forward scatterings (ie., small-$|k_x - k_x'|$ channels) are enhanced, whilst backscatterings (i.e., large momentum transfer) are suppressed exponentially.
Also, $\phi_n(y - y_0)$ is the Landau wavefunction centred at $y_0 = \ell_B^2 \,k_x$ [cf. Eq. \eqref{eq:phi-n-def}].
The integrals involve products of Hermite polynomials and are evaluated using the identity
\begin{align}
\int_{-\infty}^{\infty} d\xi \, e^{-\xi^2} H_m(\xi + z_1) \, H_n(\xi + z_2) 
= 2^n \,\sqrt{\pi} \,m! \, z_2^{n-m} \,L_m^{n-m}(-2z_1 z_2) \text{ for } m \le n,
\label{eq:hermite-identity}
\end{align}
where $L_m^{(k)}$ denotes the associated Laguerre polynomial. 
Applying this to our Landau wavefunctions, with $\xi = y/\ell_B$, $z_1 = -y_{k_x}^0/\ell_B$, $z_2 = -y_{k_x'}^0/\ell_B$, and integrating over $y$ after factoring out the Gaussian and Hermite parts, we arrive at
\begin{align}
& I^{n, n'} (k_x, k_x') = e^{i \,q_y \,\bar{y}} \, e^{-|Q|^2/4} 
\sqrt{\frac{n_< !}{n_> !}} \, \left( \frac{Q}{\sqrt{2}} \right)^{n_> - n_<} 
L_{n_<}^{n_> - n_<} \left( \frac{|Q|^2}{2} \right),
\quad n_> = \max(n,n')\,, \quad n_< = \min(n,n')\,,
\nn & \bar{y} = \frac{y_{k_x}^0 + y_{k_x'}^0} {2} \,,\quad 
\Delta y = y_{k_x'}^0 - y_{k_x}^0 = \ell_B^2(k_x' - k_x) \,,\quad 
Q = \frac{\Delta y} {\ell_B} + i\, q_y \,\ell_B\,.
\label{eq:overlap-integral}
\end{align}
The resulting matrix-element coincides with the standard LL form-factor familiar from the quantum-Hall literature, dubbed Girvin–Jach form-factor~\cite{girvin-jach}.
The Laguerre polynomial structure encodes which LLs couple under scattering, whilst the exponential factor $e^{-|Q|^2/4}$ controls how the overlap decays with orbital separation.
The formula as written applies for $n \ge n'$; for $n < n'$ replace $Q \to -Q^*$ (equivalently swap $k_x \leftrightarrow k_x'$). Since only $|I^{n,n'}|^2$ enters the matrix-element and $|Q^*| = |Q|$, this distinction is immaterial.

We use $|Q^*| = |Q|$ to obtain
\begin{align}
& \left|I^{n,n'}(k_x,k_x')\right|^2 = 
e^{-\frac{|Q|^2}{2}}\,\frac{n_<!}{n_>!} \left(\frac{|Q|^2}{2}\right)^{n_>-n_<} 
\left[
L_{n_<}^{\,n_>-n_<}\!\left(\frac{|Q|^2}{2}\right) \right]^2 
 =  \frac{n_<! \;\; e^{-\frac{\ell_B^2 \, \varsigma^2}{2}}}
{n_>!} \left( \frac{\ell_B^2\, \varsigma^2}{2}  \right )^{n_>-n_<} 
\left[ L_{n_<}^{\,n_>-n_<} \!\left( \frac{\ell_B^2 \, \varsigma^2 }{2} \right) \right]^2 ,
\nn & \varsigma = \sqrt{(k_x-k_x')^2 + q_y^2}\,.
\label{eq:overlap-squared}
\end{align}
When we multiply the Gaussian prefactor $ e^{-\boldsymbol{q}^2 \,R^2} $ with Eq.~\eqref{eq:overlap-squared}, the exponentials combine and the full exponent becomes
$ \exp\!\left[ -\left(R^2+\frac{\ell_B^2}{2}\right) \varsigma^2 \right] $.
Thus we evaluate
\begin{align} 
& \int\frac{dk_x'}{2 \, \pi}\int\frac{dq_y}{2 \, \pi} \,e^{-\, R^2\,[(k_x-k_x')^2+q_y^2]} 
\left|I^{n,n'}(k_x,k_x')\right|^2 
= \frac{n_<!}{n_>!} \, I_2\,,\quad
I_2 = \int\frac{dk_x'}{2 \, \pi} \int\frac{dq_y}{2 \, \pi} 
e^{ -A \,\varsigma^2 }\,  u^{n_>-n_<} \,\left[ L_{n_<}^{\,n_>-n_<} (u ) \right]^2,
\nn &  A=  R^2+\frac{\ell_B^2}{2}, \quad
u=\frac{\ell_B^2 \, \varsigma^2} {2} \,.
\label{eq:double-integral-setup}
\end{align}
We transform to polar coordinates, $(k_x-k_x', q_y) = \varsigma \,(\cos\phi, \sin\phi)$ and $d^2\boldsymbol{\varsigma} = \varsigma \, d\varsigma \, d\phi$. Performing the angular integral, we get
\begin{align}
I _2 &= \frac{1}{2 \, \pi} \int_0^\infty d\varsigma \, \varsigma \, e^{-A\,\varsigma^2}
 \left(\frac{\ell_B^2 \varsigma^2}{2}\right)^{n_>-n_<} 
 \left[L_{n_<}^{n_>-n_<}\left(\frac{\ell_B^2\, \varsigma^2}{2}\right)\right]^2 \nn
&= \frac{1}{4\,\pi\, A} \int_0^\infty dt \, e^{-t}\, t^{n_>-n_<} 
\left(\frac{\ell_B^2}{2A}\right)^{n_>-n_<} \left[L_{n_<}^{n_>-n_<}
\left(\frac{\ell_B^2 \,t}{2\,A}\right)\right]^2 \;
\Big (\text{on substituting } t = A\,\varsigma^2 \text{ and } 
\varsigma \, d\varsigma = \frac{dt}{2A} \Big ) \nn
&= \frac{1}{2\,\pi \,\ell_B^2} \int_0^\infty du \, e^{-\frac{2A}{\ell_B^2} u} \, 
u^{n_>-n_<} \, [L_{n_<}^{n_>-n_<}(u)]^2 \;
\Big(\text{on substituting } u = \frac{\ell_B^2 \,t}{2\,A} 
\text{ and } dt = \frac{2\,A} {\ell_B^2} du \Big ).
\end{align}
The integral is amenable to closed-form evaluation using
\begin{align}
& \int_0^\infty d \tilde x \,e^{-\, \tilde x \,\left(s+\frac{b_1+ b_2}{2}\right)}
\,\tilde x^{\tilde \mu}\, 
L_{\tilde k}^{\tilde \mu}(b_1\, \tilde x)\,L_{\tilde k}^{\tilde \mu}(b_2\, \tilde x)
= 
\frac{\Gamma(1+{\tilde \mu}+ {\tilde k})} {{\tilde k}!}\,\frac{c_2^{{\tilde k}}}
{c_0^{1+{\tilde \mu}+ {\tilde k}}} 
\,P_{\tilde k}^{({\tilde \mu},0)}\!
\left(\frac{c_1^2} { c_0\,c_2}\right),
\nn & c_0 = s + \frac{b_1+b_2}{2}\,, \quad c_2 = s - \frac{b_1+b_2}{2}, \quad 
c_1^2 = c_0 \,c_2 + 2\,b_1 \,b_2\,.
\end{align}
$P_n^{(\alpha,\beta)}({\tilde y})$ denotes the Jacobi polynomial of degree $n$, defined by 
$ P_n^{(\alpha,\beta)}({\tilde y}) = \frac{(\alpha+1)_n}{n!}\, {}_2F_1\!
\left( -n,\, n+\alpha+\beta+1;\, \alpha+1;\, \frac{1-{\tilde y}}{2} \right)$, where $(a)_n=\Gamma(a+n)/\Gamma(a)$ is the Pochhammer symbol. 
Setting $b_1 = b_2 = 1$, ${\tilde \mu} = n_> - n_<$, $ {\tilde k} = n_<$, and $ S \equiv 2\,A/\ell_B^2  $ (so that $s = S-1$ matches the exponent), the auxiliary variables evaluate to 
\begin{align}
c_0 = S \,, \quad c_2 = S-2\,, \quad c_1^2 = S\,(S-2)+2\,, \quad
S = \frac{2\, R^2} {\ell_B^2} + 1\,.
\end{align}
Note that the prefactor $n_<!/n_>!$ from Eq.~\eqref{eq:double-integral-setup} cancels exactly against the factor $\Gamma(1+{\tilde \mu}+k)/k! = \Gamma(1+n_>)/n_<! = n_>!/n_<!$ generated by the closed-form integral above. This gives
\begin{align}
&\int\frac{dk_x'}{2 \, \pi}\int\frac{dq_y}{2 \, \pi} \,\mathcal{M}^{n,n'}_{k_x,k_x'} 
= \frac{n_{\rm imp}\, V_0^2} {2 \, \pi \, \ell_B^2} \, \frac{c_2^{n_<}}{c_0^{1+n_>}} 
\, P_{n_<}^{(n_>-n_<,\,0)}\!\left(\frac{c_1^2} {c_0\,c_2}\right).
\end{align}
We observe that Eq.~\eqref{eqMpointlike} is recovered as the $R\to 0$ limit of the Gaussian result. In this short-range limit, $S=2R^2/\ell_B^2+1\to1$, $c_0\to1$, $c_2\to-1$, and $c_1^2/(c_0\,c_2)\to-1$. Using $P_{n_<}^{(n_>-n_<,0)}(-1)=(-1)^{n_<}$, 
\begin{align}
\int\frac{dk_x'}{2 \, \pi}\int\frac{dq_y}{2 \, \pi}\,\mathcal{M}^{n,n'}_{k_x,k_x'}
\;\xrightarrow{\;R\to0\;}\;
\frac{n_{\rm imp}\,V_0^2} {2 \,\pi \,\ell_B^2}\,.
\end{align}

In our calculations, we need 2 special cases:
\begin{itemize}

\item For the same LL, we have $n'= n$, implying $n_>=n_<=n$, ${\tilde \mu}=0$, and $k=n$. The general result reduces to
\begin{align}
\int\frac{dk_x'}{2 \, \pi}\int\frac{dq_y}{2 \, \pi}\,\mathcal{M}^{n,n}_{k_x,k_x'} 
= \frac{n_{\rm imp}\, V_0^2}{2 \, \pi \, \ell_B^2} \, \frac{c_2^{n}}{c_0^{n+1}}\,
P_{n}^{(0,\,0)}\!\left(\frac{c_1^2}{c_0\,c_2}\right),
\end{align}
where $P_n^{(0,0)}$ is simply the Legendre polynomial of degree $n$.

\item For adjacent LLs with $|n'-n|=1$, one has $n_> = n_<+1$ and ${\tilde \mu}=1$. Substituting directly into the general result above gives
\begin{align}
\label{eq:self-energy-final}
\int\frac{dk_x'}{2 \, \pi}\int\frac{dq_y}{2 \, \pi}\,\mathcal{M}^{n,n'}_{k_x,k_x'} 
= \frac{n_{\rm imp}\, V_0^2}{2 \, \pi \, \ell_B^2} \, 
\frac{c_2^{n_<}} {c_0^{\,n_<+2}}\,
P_{n_<}^{(1,\,0)}\!\left(\frac{c_1^2}{c_0\,c_2}\right).
\end{align}

\end{itemize}

\subsubsection{Long-ranged impurity potentials: Yukawa}
\label{sec:yukawa}

We now consider the case of a screened Coulomb interaction, modelled by a Yukawa potential. This potential arises when charged impurities are screened by a Debye-length $ \kappa^{-1}$ in the host material. We model this through a Yukawa potential of the form
\begin{align}
V(\boldsymbol{r}- \boldsymbol{r}_i) = 
 \frac{e^2}{\epsilon}\,\frac{e^{-\,\kappa \,(\boldsymbol{r}-\boldsymbol{r}_i)}} 
 {|\boldsymbol{r}-\boldsymbol{r}_i|}\,,
\label{eq:yukawa-pot}
\end{align}
where $\kappa $ is the inverse screening-length and $\epsilon$ is the effective permittivity. In the limit $ \kappa \to 0$ with fixed $V_0$, this approaches a bare Coulomb potential. In the limit $\kappa \to \infty$, the potential becomes short-ranged. The Fourier transform is obtained as
\begin{align}
u(\boldsymbol{q}) = \int d^2\boldsymbol{r} \, 
e^{i \, \boldsymbol{q} \cdot \boldsymbol{r}} \, V(\boldsymbol{r}) 
= \frac{2\,\pi\, e^2} {\epsilon \, \sqrt{ {\boldsymbol q}^2 + \kappa^2}} \,,
\end{align} 
leading to
\begin{align}
u(\boldsymbol q) \,u(-\boldsymbol q) =\frac{ {\tilde V_0}^2} { {\boldsymbol q}^2 + \kappa^2 } 
\,, \quad  {\tilde V_0}^2 =  \frac{4\,\pi^2\, e^4} {\epsilon^2}\,.
\end{align}
Unlike the Gaussian potential, the Yukawa interaction depends only on the magnitude of the momentum transfer. Consequently, the disorder matrix-element may be written directly in terms of the LL form-factor, 
\begin{align} 
& M_{n,n'} = n_{\rm imp} 
\int\frac{d^2q}{(2\,\pi)^2} 
\frac{ {\tilde V_0}^2 }{ \kappa^2+{\boldsymbol q}^2 } \;  \frac{e^{-\,\frac{\ell_B^2 \,{\boldsymbol q}^2} {2} } \, \, n_<!} {n_>!} 
\left(\frac{\ell_B^2 \,{\boldsymbol q}^2}{2}\right)^{n_>-n_<} 
\left[ L_{n_<}^{\,n_>-n_<} \! \left(\frac{\ell_B^2 \,{\boldsymbol q}^2}{2}\right) \right]^2, 
\quad n_>=\max(n,n'), \quad n_<=\min(n,n')\,.
\end{align} 
 Since the integrand depends only on the magnitude of the momentum transfer, we transform to polar coordinates.
 The matrix-element therefore becomes 
 \begin{align}  
 M_{n,n'} = \frac{n_{\rm imp}\, {\tilde V_0}^2}{2 \, \pi} 
 \frac{n_<!}{n_>!} \int_0^\infty dq\, 
 \frac{ q\,e^{-\,\frac{\ell_B^2 \,{\boldsymbol q}^2} {2} } }    { \kappa^2+{\boldsymbol q}^2 } 
 \left(\frac{\ell_B^2 \,{\boldsymbol q}^2}{2}\right)^{n_>-n_<} \left[ L_{n_<}^{\,n_>-n_<} \!
 \left(\frac{\ell_B^2 \,{\boldsymbol q}^2}{2}\right) \right]^2. 
 \end{align}  
Using $ u =\frac{\ell_B^2 \,{\boldsymbol q}^2}{2}$ and $ q\,dq =\frac{du} {\ell_B^2}$, we obtain 
 \begin{align} 
 M_{n,n'} = \frac{n_{\rm imp}\, {\tilde V_0}^2} {4\,\pi} \,\frac{n_<!}{n_>!} 
 \int_0^\infty du \, \frac{ e^{-u} \,u^{\,n_>-n_<} 
 \left[ L_{n_<}^{\,n_>-n_<}(u) \right]^2 } { u +\beta}\,, \quad
\beta = \frac{\kappa^2 \,\ell_B^2}{2}\,. 
 \end{align} 
The two cases required in practice are as follows, both evaluated numerically.
For the same LL, $n'=n$ (using $  n_>=n_<=n$, $n_<!/n_>!=1$):
\begin{align}
\label{eq:yukawa-nn}
M_{n,n} = \frac{n_{\rm imp} \,{\tilde V_0}^2} {4 \, \pi}
\int_0^\infty du\,\frac{e^{-u}\left[L^0_n(u)\right]^2}{u+\beta}\,.
\end{align}
For adjacent LLs, $|n'-n|=1$ [using $  n_>=n_<+1$, $n_<!/n_>!=1/(n_<+1)$]:
\begin{align}
\label{eq:yukawa-adj}
M_{n,n\pm1} = \frac{n_{\rm imp}\, {\tilde V_0}^2} {4\, \pi \, (n_<+1)}
\int_0^\infty du\,\frac{u\,e^{-u}\left[L_{n_<}^{\,1}(u)\right]^2}{u+\beta}\,.
\end{align}

\section{Selection rules in the ultraquantum limit}
\label{sec:selrules-ql}

In the quantum limit, an LLL is partially occupied because $\upmu$ cuts the middle of the disorder-broadened LLL(s) --- states below (above) the chemical potential are filled (empty).

Only the LLs closest in energy to the partially occupied LLLs need be retained in the low-energy conductivity. The Dirac case is very simple and has been relegated to Appendix~\ref{appdirac}. On the other hand, the LLs of a GNR exhibit a non-monotonic stretched-checkmark pattern, with descending and ascending level branches \cite{barati-nlsm-qhe}, making the bookkeeping harder. Hence, we proceed to elaborate on the selection rules in this section. Let us define
\begin{align}
\Delta_- = \begin{cases} 
E_{n_g-1} - E_g &\text{ for } \rho \neq 2\, N \\
E_{n_1-1} - E_g &\text{ for } \rho = 2\, N
\end{cases} \text{   and   }
\Delta_+ = \begin{cases} 
E_{n_g+1} - E_g &\text{ for } \rho \neq 2\, N \\
E_{n_1 + 2} - E_g &\text{ for } \rho = 2\, N
\end{cases} .
\end{align}
For the non-degenerate LLL case, we have denoted the element of $\lbrace n_g \rbrace$ as $n_g$ itself, while $\lbrace n_g \rbrace = \lbrace n_1, n_2 \rbrace$ for $\rho = 2\, N$.
The spacing between $E_g$ and $-\, E_g$  is obviously $2\,E_{g}$, which, within the ultraquantum-limit approximation $\upmu\simeq E_g$ adopted throughout, is $2\,E_{g} \simeq 2\, \upmu $. Since states with negative energy may have to be included, the matrix-element must be generalised to contain the $s=\pm$ index, such that
\begin{align}
\label{eqmat}
\mathcal{M}_{k_x, \tilde k_x}^{n,s; \tilde n, \tilde s} = 
\overline{ \langle \tilde k_x,  \tilde n, \tilde s| V(\boldsymbol{r}) \, |k_x, n, s \rangle \, 
 \langle k_x, n, s| V (\boldsymbol{\tilde r}) \, |\tilde k_x,  \tilde n, \tilde s \rangle } \,.
\end{align}
The index $s$ equals $+$ ($-$) depending on whether the closest-to-$E_g$ level has positive (negative) energy. For each regime below, we evaluate the relevant self-energies using the ultraquantum-limit approximation: assume positive chemical potential $\upmu > 0$ close to the LLL bottom ($\upmu \simeq E_g$), truncate the sum to dominant contributions (single term $n_0 =n_g$ or $n_0 \in \lbrace n_1, n_2 = n_1+1 \rbrace $ for degenerate cases), and evaluate self-energies by using Eqs.~\eqref{eqsel_LLL} and \eqref{eqsel_neigh}.

\subsection{Nonzero matrix-elements}

From the expressions in Sec.~\ref{sec:model-ham}, for the GNR, we have
\begin{align}
j_x  =  \sqrt{\frac{e^2\, \varepsilon_c}{2\,m^*}}\; (a+a^\dagger)\, \sigma_x \,,\quad
j_y  =  i\,\sqrt{\frac{e^2\, \varepsilon_c}{2\,m^*}}\; (a-a^\dagger)\, \sigma_x \,,
\end{align}
\begin{align}
\label{eqmat0}
& \langle \tilde{k}_x,\tilde{n},\tilde{s}| \,v_x \, |k_x,n,s\rangle
= \frac{1}{L_x}\int dx\, e^{i(k_x-\tilde k_x) \,x}\;
\sqrt{\frac{\varepsilon_c}{2\,m^*}}\,
\langle\phi_{\tilde{n}}|\,(a+a^\dagger) \,|\phi_n\rangle\,
\langle \tilde{s}|\sigma_x|s\rangle
\nn & = \delta_{k_x,\tilde k_x}\,
\sqrt{\frac{\varepsilon_c}{2\,m^*}}\,
\left ( \sqrt{n}\; \delta_{\tilde{n},n-1}+\sqrt{n+1}\; \delta_{\tilde{n},n+1} \right)
\langle \tilde{s}|\sigma_x|s\rangle\,,
\nn & \langle \tilde{k}_x,\tilde{n},\tilde{s}| \,v_y \, |k_x,n,s\rangle
= \frac{i}{L_x}\int dx\, e^{i(k_x-\tilde k_x) \,x}\;
\sqrt{\frac{\varepsilon_c}{2\,m^*}}\,
\langle\phi_{\tilde{n}}|\,(a-a^\dagger) \,|\phi_n\rangle\,
\langle \tilde{s}|\sigma_x|s\rangle
\nn & = i\,\delta_{k_x,\tilde k_x}\,
\sqrt{\frac{\varepsilon_c}{2\,m^*}}\,
\left ( \sqrt{n}\; \delta_{\tilde{n},n-1}-\sqrt{n+1}\; \delta_{\tilde{n},n+1} \right)
\langle \tilde{s}|\sigma_x|s\rangle\,,
\end{align}
where the $1/L_x$ prefactor from the two plane-wave normalizations combines with $\int dx\, e^{i\,(k_x-\tilde k_x) \,x}$ to collapse onto $\delta_{k_x,\tilde k_x}$.\footnote{This follows because each wavefunction now has a plane-wave factor $1/\sqrt{L_x}$. The product gives $1/L_x$, and the integral over $x$ gives $L_x$ times the Kronecker delta, thus canceling the $1/L_x$.}
The oscillator factor is nonzero only for $|\tilde{n}-n|=1$ for both $v_x$ and $v_y$. The pseudospin factor is generically nonzero and imposes no further restriction. Therefore,
\begin{align}
|\langle k_x,n,s|\,v_x\, |\tilde{k}_x,\tilde{n},\tilde{s}\rangle|^2 \neq 0\,,\quad
\langle k_x,n,s|\,v_y\,|\tilde{k}_x,\tilde{n},\tilde{s}\rangle\,\langle \tilde{k}_x,\tilde{n},\tilde{s}|\,v_x\,|k_x,n,s\rangle \neq 0
\quad\Longleftrightarrow\quad |\tilde{n}-n|=1\,.
\end{align}
This kills many terms in the lists given in the analysis below. We will call it vv selection rule.

Now we evaluate $|\langle k_x, n, s\,|\,v_\mu\,|\tilde{k}_x,\tilde{n},\tilde{s}\rangle|^2$ explicitly for $|\tilde{n}-n|=1$.
For the pseudospin factor, we get
\begin{align}
\langle s|\sigma_x|\tilde{s}\rangle 
&= \frac{1}{2}\Big[ \tilde s\,\mathrm{sgn}(\sin\theta_{\tilde n})\,A_n^{s+}A_{\tilde{n}}^{\tilde{s}-}
+ s\,\mathrm{sgn}(\sin\theta_n)\,A_n^{s-}A_{\tilde{n}}^{\tilde{s}+} \Big]
\nn &= \frac{1}{2}\Big[
\tilde s\,\mathrm{sgn}(\sin\theta_{\tilde n})\sqrt{(1+s\cos\theta_n)(1-\tilde s\cos\theta_{\tilde n})}
+ s\,\mathrm{sgn}(\sin\theta_n)\sqrt{(1-s\cos\theta_n)(1+\tilde s\cos\theta_{\tilde n})} \Big],
\end{align}
since $\sigma_x=\begin{bmatrix}0&1\\1&0\end{bmatrix}$ swaps the two spinor components, so the bra's "$+$" amplitude pairs with the ket's "$-$" amplitude and vice versa.

We need
\begin{align}
& |\langle n, s|\sigma_x|\tilde n, \tilde s\rangle|^2
\nn & = \frac{1}{4}\Big[(1+s\cos\theta_n)(1-\tilde s\cos\theta_{\tilde n})
+(1-s\cos\theta_n)(1+\tilde s\cos\theta_{\tilde n})\Big]
+ \frac{1}{2}\,s\,\tilde s\,\mathrm{sgn}(\sin\theta_n\sin\theta_{\tilde n})
\sqrt{(1-s^2\cos^2\theta_n)\,(1-\tilde s^2\cos^2\theta_{\tilde n})} \nn
&= \frac{1}{2}\Big[1 - s\,\tilde s\,\cos\theta_n\cos\theta_{\tilde n}\Big]
+ \frac{1}{2}\,s\,\tilde s\,\sin\theta_n\sin\theta_{\tilde n}
\nn
&= \frac{1}{2}\Big[1 - s\,\tilde s\,\cos(\theta_n+\theta_{\tilde n})\Big] ,
\label{eq:sigmax_sq}
\end{align}
where in the last-but-one step we used
$\mathrm{sgn}(\sin\theta_n\sin\theta_{\tilde n})|\sin\theta_n||\sin\theta_{\tilde n}|
=\sin\theta_n\sin\theta_{\tilde n}$.

Combining both factors, the squared matrix
element for the two cases $\tilde n = n\pm1$ reads
\begin{align}
|\langle k_x,n,s\,|\,v_x\,|\,\tilde k_x,\tilde n,\tilde s\rangle|^2
&= \frac{\varepsilon_c}{2 \, m^*}\,\delta_{k_x,\tilde k_x}\times
\begin{cases}
\dfrac{(n+1)}{2}\big[1-s  \,\tilde s\cos(\theta_n+\theta_{n+1})\big] & \text{ for } \tilde n = n+1 \\
\dfrac{n}{2}\big[1-s \,\tilde s\cos(\theta_n+\theta_{n-1})\big] & \text{ for } \tilde n = n-1
\end{cases} \nn
&= \frac{\varepsilon_c \, \max(n,\tilde n)}  {4 \, m^*}\,\delta_{k_x,\tilde k_x}\,
\left[1 - s\,\tilde s\,
\frac{\Delta^2 - (\varepsilon_n-\varepsilon_r)\,(\varepsilon_{\tilde n}-\varepsilon_r)}
{E_n \,E_{\tilde n}}\right] ,
\label{eq:vx_matrix_element}
\end{align}
where we used $\cos\theta_n = \Delta/E_n$, $\sin\theta_n = (\varepsilon_n-\varepsilon_r)/E_n$, and
$\cos(\theta_n+\theta_{\tilde n})=\cos\theta_n\cos\theta_{\tilde n}-\sin\theta_n\sin\theta_{\tilde n}$.
As a check, for $n=\tilde n$ this gives $\tfrac12[1-s\tilde s\cos2\theta_n]$; these sum to $1$, as required by $\sigma_x^2=\mathbb{I}_{2\times 2}$. Neither branch vanishes identically: $\sigma_x$ does not commute with $H_n$, so orthogonal eigenstates of $H_n$ need not be annihilated by $\sigma_x$.

For the Hall conductivity, the relevant quantity is the cross term
$\langle k_x,n,s|\,v_y\,|\tilde k_x,\tilde n,\tilde s\rangle\,
\langle \tilde k_x,\tilde n,\tilde s|\,v_x\,|k_x,n,s\rangle$,
which appears in the Kubo--Bastin expression for $\sigma_{xy}^{ql}$.
Using Eq.~\eqref{eqmat0}, the two ladder branches give
\begin{align}
\langle k_x,n,s|\,v_y\,|\tilde k_x,\tilde n,\tilde s\rangle\,
\langle \tilde k_x,\tilde n,\tilde s|\,v_x\,|k_x,n,s\rangle
&= \frac{\varepsilon_c}{2\,m^*}\,\delta_{k_x,\tilde k_x}\times
\begin{cases}
i\,\dfrac{(n+1)}{2}\big[1-s\,\tilde s\cos(\theta_n+\theta_{n+1})\big]
& \text{for } \tilde n=n+1,\\[6pt]
-\,i\,\dfrac{n}{2}\big[1-s\,\tilde s\cos(\theta_n+\theta_{n-1})\big]
& \text{for } \tilde n=n-1,
\end{cases} \nn
&= i\,\mathrm{sgn}(\tilde n-n)\,
\frac{\varepsilon_c\,\max(n,\tilde n)}{4\,m^*}\,
\delta_{k_x,\tilde k_x} \left[ 1-s\,\tilde s\,
\frac{\Delta^2-(\varepsilon_n-\varepsilon_r)
(\varepsilon_{\tilde n}-\varepsilon_r)} {E_n \, E_{\tilde n}} \right] \nn
&= i\,\mathrm{sgn}(\tilde n-n)\,
|\langle k_x,n,s\,|\,v_x\,|\,\tilde k_x,\tilde n,\tilde s\rangle|^2 \,.
\label{eq:vyvx_matrix_element}
\end{align}

It is worth spelling out the usual Landau-degeneracy count here. With periodic boundary conditions implemented along the $x$-direction, $k_x = 2\,\pi\, j/L_x$ for integer $j$, and the corresponding guiding centre $y_0 = \ell_B^2\, k_x$ must lie within the physical extent $L_y$ of the sample. The number of allowed values of $j$ (equivalently, of $k_x$) is therefore
\begin{align}
N_\phi = \frac{L_y/\ell_B^2}{2\,\pi/L_x} = \frac{L_xL_y}{2\,\pi\,\ell_B^2} 
= \frac{\mathcal V}{2\,\pi\,\ell_B^2}\,,
\end{align}
It represents the magnetic flux through the sample in units of the flux quantum. Crucially, this depends only on the product $\mathcal V = L_x \, L_y$. The $k_x$ sum therefore contributes a factor of $\mathcal V/(2\,\pi\,\ell_B^2)$ (equivalently $L_x \, L_y/(2\,\pi\,\ell_B^2)$) upon integration, leading to
\begin{align}
\label{eqdotsq}
\sum_{k_x,\tilde k_x}|\langle k_x,n,s\,|\,v_x\,|\,\tilde k_x,\tilde n,\tilde s\rangle|^2
&= \sum_{k_x}\frac{\varepsilon_c\,\max(n,\tilde n)}{4\,m^*} \left[ 1-s\,\tilde s\,
\frac{\Delta^2-(\varepsilon_n-\varepsilon_r)(\varepsilon_{\tilde n}-\varepsilon_r)}{E_n \, E_{\tilde n}}\right]\nn
&= \frac{\mathcal V\,e^2B^2\,\max(n,\tilde n)}
{8\pi(m^*)^2} \left[
1-s\,\tilde s\,\frac{\Delta^2-(\varepsilon_n-\varepsilon_r)
(\varepsilon_{\tilde n}-\varepsilon_r)}
{E_n \, E_{\tilde n}}\right].
\end{align}

\subsection{Expressions for conductivity}
\label{sec-cond-gnr}

Starting from Eq.~\eqref{eq:sig-general}, remembering that $\mathrm{Re}[\Sigma^R_n(\upmu)] = 0 $, the longitudinal dc conductivity in the ultraquantum limit is captured by
\begin{align}
\label{eq:sigxx-ql}
\sigma_{xx}^{ql} (\upmu) &= 
 \sum_{n_0 \in \lbrace n_{g} \rbrace} \sum_{n \in \lbrace n_{gn} \rbrace} 
\tilde \sigma_{xx}^{n_0; n, s} (\upmu)\,, 
 \nn  \tilde \sigma_{xx}^{n_0; n, s} (\upmu) & =  
\sum_{k_x, \,k_x'} \frac{e^2 }{8\, \pi \, \mathcal{V}} \;
|\langle \tilde k_x,  n_0, +|\; v_x \; |k_x, n, s \rangle|^2
\,\mathcal{A}_{n_0, +}( \upmu) \, \mathcal{A}_{n, s}( \upmu)
\nn & = \quad 
\frac{e^4\,B^2\,\max(n_0,n)}{64\,\pi^2\,(m^*)^2}
\left[1-s\,\frac{\Delta^2-(\varepsilon_n-\varepsilon_r)(\varepsilon_{n_0}-\varepsilon_r)}{E_n\,E_{n_0}}\right]
\mathcal{A}_{n_0, +}( \upmu) \, \mathcal{A}_{n, s}( \upmu)\,,
\nn  \mathcal{A}_{n, s} ( \upmu) & = \frac{2 \,\Gamma_{n, s}(\upmu)}
{ (\upmu-s\, E_n)^2+\Gamma_{n, s}^2(\upmu) }\,,\quad
\Gamma_{n, s}(\upmu) =  - \,\mathrm{Im}[\Sigma_{n,s}(\upmu) ] \,;
\end{align}
We note that, since $D_{n_0,+}(\upmu)\simeq0$ in the ultraquantum limit,
$\mathcal B_{n_0,+}(\upmu)\simeq0$, and the
$\mathcal B_{n_0,+}(\upmu)\,\mathcal B_{n,s}(\upmu)$
term in Eq.~\eqref{eq:sig-general} can be neglected.

Similarly, the Hall conductivity is
\begin{align}
& \sigma_{xy}^{ql}(\upmu) = \sum_{n_0\in\{n_g\}}
\sum_{n\in\{n_{gn}\}}\,\mathrm{sgn}(n-n_0)\;
\tilde\sigma_{xy}^{n_0;n,s}(\upmu)\,,
\nn &
\tilde\sigma^{xy}_{n_0;n,s}(\upmu) = \sum_{k_x,\,k_x'}\frac{ e^2} {4\,\pi\,\mathcal V}\,
\big|\langle \tilde k_x,n_0,+|\,v_x\,|k_x,n,s\rangle\big|^2\,
\Big[\mathcal B_{n,s}(\upmu)\,\mathcal A_{n_0,+}(\upmu) 
- \mathcal A_{n,s}(\upmu)\,\mathcal B_{n_0,+}(\upmu)
\Big]\,, \nn &
\mathcal B_{n,s}(\upmu) = 
\frac{2 \, (\upmu-s\, E_n)}{(\upmu-s\, E_n)^2+\Gamma^2_{n,s}(\upmu) }\,.
\label{eq:sigxy-ql}
\end{align}
Since the contributions from the ordered pairs $(n_0,n)$ and $(n,n_0)$ are identical, the double sum in Eq.~\eqref{eq:sig-general} has been reduced to a single sum over distinct $n_g$--$n_{gn}$ pairs, resulting in the prefactor $e^2/(4\pi\mathcal V)$. In the above expressions, the superscript ``$ql$'' denotes the quantum limit. Both the SCBA self-energy of the LLL [analogous to Eq.~\eqref{eqsel_LLL}], and the self-energy of its neighbours [analogous to Eq.~\eqref{eqsel_neigh}], are already purely imaginary at the level of the ultraquantum-limit truncation used throughout this work (since in both the cases the truncation $\upmu-E_g -\Sigma_n \to-\Sigma_n $ discards $\mathrm{Re}\,\Sigma_n $ identically). Consequently $D_{n_g,+}(\upmu)\approx0$ and $\mathcal B_{n_g,+}(\upmu)\approx0$, so the LLL's dispersive kernel drops out of Eq.~\eqref{eq:sigxy-ql} entirely, leaving
\begin{align}
\sigma_{xy}^{ql}(\upmu) &= 
\frac{e^2}{4\pi\mathcal V}\sum_{n_0\in\{n_g\}}\sum_{n\in\{n_{gn}\}}\mathrm{sgn}(n-n_0)\,\sum_{k_x,\,k_x'}
\big|\langle \tilde k_x,n_0,+|v_x|k_x,n,s\rangle\big|^2\,\mathcal A_{n_0,+}(\upmu)\,\mathcal B_{n,s}(\upmu)
\nn & 
= \frac{e^4\,B^2}{32\,\pi^2\,(m^*)^2}\sum_{n_0\in\{n_g\}}\sum_{n\in\{n_{gn}\}}\mathrm{sgn}(n-n_0)\,\max(n_0,n)\,
\left[1-s\,\frac{\Delta^2-(\varepsilon_n-\varepsilon_r)(\varepsilon_{n_0}-\varepsilon_r)}{E_n\,E_{n_0}}\right]
\mathcal A_{n_0,+}(\upmu)\,\mathcal B_{n,s}(\upmu)\,.
\label{eq:sigxy-ql-final}
\end{align}
This conclusion extends unchanged to the doubly-degenerate LLL case, Eq.~\eqref{eqsiglll-deg}: any self-energy equation produced by the $\upmu \simeq E_g$ truncation takes the schematic form $\Sigma_a = \sum_b M_{ab}/(-\Sigma_b)$ with real coefficients $M_{ab}$. Substituting $\Sigma_b=i\, \Gamma_b$ converts any such equation into a manifestly real equation for the $\Gamma_b$. Consequently $\mathrm{Re}[\Sigma_{n,s}(\upmu)]=0$ throughout this truncation scheme --- for the LLL (degenerate or not) and for its neighbours alike. Hence, no separate calculation of self-energy is required for Eq.~\eqref{eq:sigxy-ql-final}.

\begin{figure}[t!]
    \centering
 \includegraphics[width= 0.85 \linewidth]{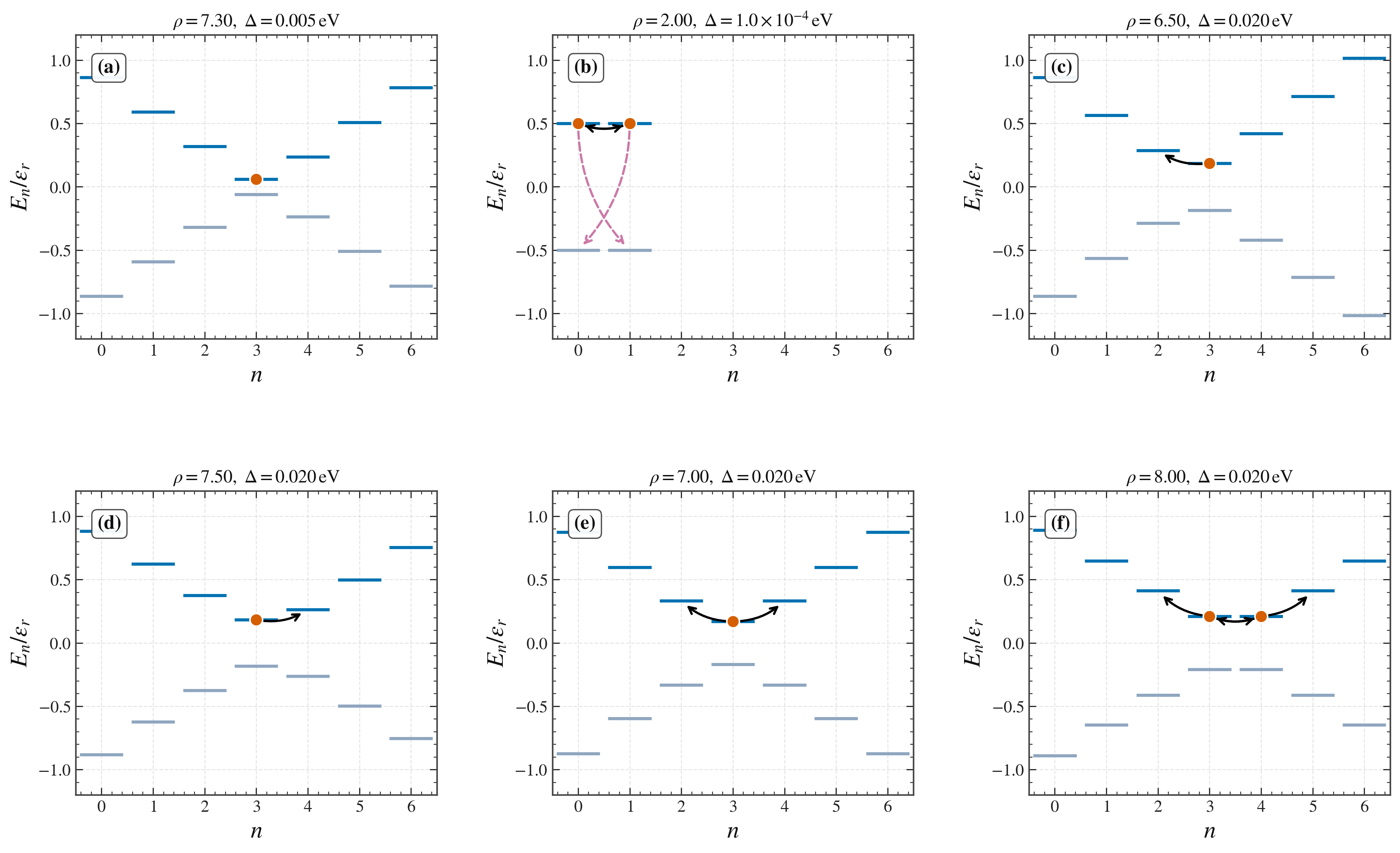}
    \caption{Schematics illustrating the selection rules of Secs.~\ref{sec-cond-gnr1}---\ref{sec-cond-gnr4}. The staircase patterns in each subfigure displays the LL spectrum as a function of $n$, with the orange dot indicating the LLL(s).
\textbf{(a)} Case 1 of Sec.~\ref{sec-cond-gnr1}: vv selection rule kills all terms.
\textbf{(b)} Case 2 of Sec.~\ref{sec-cond-gnr1}: pairing among the doubly-degenerate LLLs and with the neighbouring LLs (in the hole-branch) are shown via solid black (intraband, $+ \leftrightarrow +$) and dashed purple(interband, $+ \leftrightarrow -$) arrows, respectively.
\textbf{(c)} Case 1 of Sec.~\ref{sec-cond-gnr2}: neighboring LL is $n_g-1$, indicated by the black arrow.
 \textbf{(d)} Case 2 of Sec.~\ref{sec-cond-gnr2}: neighboring LL is $n_g+1$, indicated by the black arrow.
\textbf{(e)} Sec.~\ref{sec-cond-gnr3}: neighbouring LLs $n_g-1$ and $n_g+1$ are degenerate in energy and both contribute, indicated by black arrows.
\textbf{(f)} Sec.~\ref{sec-cond-gnr4}:  LLL itself is doubly-degenerate comprising indices $n_1$ and $(n_1+1) $, and black arrows indicate pairing with the neighbouring LLs $\in \lbrace n_1-1, \,n_1+2 \rbrace $ (also with degenerate energies).
\label{fig:gnr-staircase-cases}}
\end{figure}

Below we show all the possible cases and resulting final contribution to $\sigma_{xy}^{ql}(\upmu)$.
Fig.~\ref{fig:gnr-staircase-cases} summarises the results schematically.

\subsubsection{$2 \,E_g < \min\!\left\{ \Delta_+,\, \Delta_-\right\} $}
\label{sec-cond-gnr1}

The contribution from the LL with energy $(-\,E_{n_g})$ must be retained whenever $2 \,E_g < \min\!\left\{ \Delta_+,\, \Delta_-\right\} $. Here $s=-$ for the closest-to-$E_g$-level index, as it involves the negative-energy axis.
\begin{itemize}

\item {Case 1: $\rho \neq 2\, N $}
\\Spectral-weight overlaps with LLL to compute conductivity (2 terms in the sum for $\sigma_{xx}^{ql}$): $\lbrace n_g, + \rbrace \to \lbrace n_g, + \rbrace $ and $\lbrace n_g, + \rbrace \to \lbrace n_g, - \rbrace $. The complete set of states and LLs comprise $ \lbrace \psi_{n_g, \pm} \rbrace $ and $ \lbrace \pm E_{n_g} \rbrace$, leading to
\begin{align}
\sigma_{\mu \nu}^{ql} = \tilde \sigma_{\mu \nu}^{n_g; n_g, +} + \tilde \sigma_{\mu \nu}^{n_g; n_g, -} \,.
\end{align}

We note that the self-energy of the LLL is obtained by truncating the sum to the dominant term and evaluating at $\upmu \simeq E_g$:
\begin{align}
\label{eqsiglll-nondeg}
\Sigma_{n_g,+}(\upmu) & \approx
 L_x \int \frac{ dk_x'} {2 \, \pi}  \;
\frac{ \mathcal{M}_{k_x, k_x'}^{n_g,+; n_g,+ } } {\upmu -  E_g - \Sigma_{n_g, +} (\upmu)}
 \approx
 L_x \int \frac{ dk_x'} {2 \, \pi}  \;
\frac{ \mathcal{M}_{k_x, k_x'}^{n_g,+; n_g,+ } } { -\, \Sigma_{n_g, +} (\upmu)}\,.
\end{align}
From the above equation, we solve for $\Sigma_{n_g,+}(\upmu)$ self-consistently (SCBA).
We also need
\begin{align}
\Sigma_{n_g, -}(\upmu) & \approx
 L_x \int \frac{ dk_x'} {2 \, \pi}  \;
\frac{ \mathcal{M}_{k_x, k_x'}^{ n_g,-; n_g,+ } }
{\upmu -  E_g - \Sigma_{n_g, +} (\upmu)} \approx
 L_x \int \frac{ dk_x'} {2 \, \pi}  \;
\frac{ \mathcal{M}_{k_x, k_x'}^{ n_g,-; n_g,+ } } { - \,\Sigma_{n_g, +} (\upmu)}  \,.
\end{align}
However, after combining with the vv selection rule, $\sigma_{\mu \nu}^{ql} = 0$.

\item {Case 2: $\rho = 2\, N $}
\\Spectral-weight overlaps with doubly-degenerate LLLs (8 terms in the sum for $\sigma_{xx}^{ql}$): $\lbrace n_1, + \rbrace \to \lbrace n_1, + \rbrace $, $\lbrace n_1, + \rbrace \to \lbrace n_2, + \rbrace $ , $\lbrace n_2, + \rbrace \to \lbrace n_2, + \rbrace $, $\lbrace n_2, + \rbrace \to \lbrace n_1, + \rbrace $, $\lbrace n_1, + \rbrace \to \lbrace n_1, - \rbrace $, $\lbrace n_1, + \rbrace \to \lbrace n_2, - \rbrace $, $\lbrace n_2, + \rbrace \to \lbrace n_1, - \rbrace $, $\lbrace n_2, + \rbrace \to 
\lbrace n_2, - \rbrace $. The complete sets comprise $ \lbrace \psi_{n_1, \pm} , \psi_{n_2, \pm}\rbrace$ and $ \lbrace \pm E_{n_1} =\pm E_{n_2} \rbrace$, leading to
\begin{align}
\sigma_{\mu \nu}^{ql} = \tilde \sigma_{\mu \nu}^{n_1; n_1, +} + \tilde \sigma_{\mu \nu}^{n_1; n_2, +} +\tilde \sigma_{\mu \nu}^{n_2; n_1, +} + \tilde \sigma_{\mu \nu}^{n_2; n_2, +}
+ \tilde \sigma_{\mu \nu}^{n_1; n_1, -} + \tilde \sigma_{\mu \nu}^{n_1; n_2, -} + \tilde \sigma_{\mu \nu}^{n_2; n_1, -} + \tilde \sigma_{\mu \nu}^{n_2; n_2, -} \,.
\end{align}

For the degenerate LLL case, the self-energies of both LL indices must be computed self-consistently:
\begin{align}
\label{eqsiglll-deg}
\Sigma_{n_1,+}(\upmu) & \approx
 L_x \int \frac{ dk_x'} {2 \, \pi}  \;
\frac{ \mathcal{M}_{k_x, k_x'}^{n_1,+; n_1,+ } }
{\upmu -  E_g - \Sigma_{n_1, +} (\upmu)}
+
L_x \int \frac{ dk_x'} {2 \, \pi}  \;
\frac{ \mathcal{M}_{k_x, k_x'}^{n_1, +;n_1+1,+ } }
{\upmu -  E_g - \Sigma_{n_1+1, +} (\upmu)} \nn
& \approx
 L_x \int \frac{ dk_x'} {2 \, \pi}  \;
\frac{ \mathcal{M}_{k_x, k_x'}^{n_1,+; n_1,+ } } { - \, \Sigma_{n_1, +} (\upmu)}
+
L_x \int \frac{ dk_x'} {2 \, \pi}  \;
\frac{ \mathcal{M}_{k_x, k_x'}^{n_1, +;n_1+1,+ } } { -\, \Sigma_{n_1+1, +} (\upmu)}\,, \nn
\Sigma_{n_2,+}(\upmu) & \approx
 L_x \int \frac{ dk_x'} {2 \, \pi}  \;
\frac{ \mathcal{M}_{k_x, k_x'}^{n_1+1, +; n_1, + } }
{\upmu -  E_g - \Sigma_{ n_1} (\upmu)}
+
L_x \int \frac{ dk_x'} {2 \, \pi}  \;
\frac{ \mathcal{M}_{k_x, k_x'}^{n_1+1,+; n_1 + 1, + } }
{\upmu -  E_g - \Sigma_{ n_1+1} (\upmu)} \nn
& \approx L_x \int \frac{ dk_x'} {2 \, \pi}  \;
\frac{ \mathcal{M}_{k_x, k_x'}^{n_1+1, +; n_1, + } } { - \,\Sigma_{ n_1, +} (\upmu)}
+
L_x \int \frac{ dk_x'} {2 \, \pi}  \;
\frac{ \mathcal{M}_{k_x, k_x'}^{n_1+1, +; n_1+1,+ } } {- \,\Sigma_{ n_1+1, +} (\upmu)} \,, 
\end{align}

By solving the above 2 equations self-consistently, we compute $\Sigma_{n_1,+}(\upmu)$ and $\Sigma_{n_2,+}(\upmu)$. Furthermore, we need
\begin{align}
\Sigma_{n_1, -}(\upmu) & \approx
 L_x \int \frac{ dk_x'} {2 \, \pi}  \;
\frac{ \mathcal{M}_{k_x, k_x'}^{n_1,-; n_1,+ } } {-\, \Sigma_{ n_1, +} (\upmu)}
+
L_x \int \frac{ dk_x'} {2 \, \pi}  \;
\frac{ \mathcal{M}_{k_x, k_x'}^{n_1, -; n_1+1, + } }{ - \,\Sigma_{ n_1+1, + } (\upmu)}\,,\nn
\Sigma_{n_2, -}(\upmu) & \approx
 L_x \int \frac{ dk_x'} {2 \, \pi}  \;
\frac{ \mathcal{M}_{k_x, k_x'}^{n_1+1, -; n_1, + } } { -\, \Sigma_{ n_1, +} (\upmu)}
+
L_x \int \frac{ dk_x'} {2 \, \pi}  \;
\frac{ \mathcal{M}_{k_x, k_x'}^{n_1 +1, -;  n_1+1, + }} { -\, \Sigma_{n_1+1, +} (\upmu)}\,.
\end{align}

\end{itemize}
After combining with the vv selection rule, 
\begin{align}
\sigma_{\mu \nu}^{ql} =  \tilde \sigma_{\mu \nu}^{n_1; n_1+1, +} +\tilde \sigma_{\mu \nu}^{n_1+1; n_1, +} 
+ \tilde \sigma_{\mu \nu}^{n_1; n_1+1, -} + \tilde \sigma_{\mu \nu}^{n_1+1; n_1, -}  \,.
\end{align}

\subsubsection{$2 \,E_g > \min\!\left\{ \Delta_+,\, \Delta_-\right\} $ and $ \rho \neq \mathbb{Z}^+ $}
\label{sec-cond-gnr2}

There is no degeneracy of any LLs and only positive-energy LLs (i.e., $s = \tilde s = + $) are to be considered.
\begin{itemize}

\item {Case 1: $\Delta_- < \Delta_+$}
\\Spectral-weight overlaps with LLL to compute conductivity (2 terms in the sum for $\sigma_{xx}^{ql}$): $n_g \to n_g $ and $n_g \to n_g-1$. The complete sets comprise $ \lbrace \psi_{n_g, +}, \,\psi_{n_g-1, +} \rbrace $ and $ \lbrace  E_{g} , \, E_{n_g-1}\rbrace$, leading to
\begin{align}
\sigma_{\mu \nu}^{ql} = \tilde \sigma_{\mu \nu}^{n_g ; n_g, +} + \tilde \sigma_{\mu \nu}^{n_g; n_g-1, +}  \,.
\end{align}
We use the LLL self-energy from Eq.~\eqref{eqsiglll-nondeg} and the neighboring LL self-energy:
\begin{align}
\label{eqsig-delm}
\Sigma_{n_g-1, +}(\upmu) &  \approx
 L_x \int \frac{ dk_x'} {2 \, \pi}  \;
\frac{ \mathcal{M}_{k_x, k_x'}^{ n_g-1,+; n_g,+ } } { - \,\Sigma_{n_g, +} (\upmu)}  \,.
\end{align}
After combining with the vv selection rule, 
\begin{align}
\sigma_{\mu \nu}^{ql} =   \tilde \sigma_{\mu \nu}^{n_g; n_g-1, +}  \,.
\end{align}

\item {Case 2: $\Delta_- > \Delta_+$}
\\Spectral-weight overlaps with LLL to compute conductivity (2 terms in the sum for $\sigma_{xx}^{ql}$): $n_g \to n_g $ and $n_g \to n_g+1$. The complete sets comprise $ \lbrace \psi_{n_g, +}, \,\psi_{n_g+1, +} \rbrace $ and $ \lbrace  E_{g} , \, E_{n_g+1}\rbrace$, leading to
\begin{align}
\sigma_{\mu \nu}^{ql} = \tilde \sigma_{\mu \nu}^{n_g ; n_g, +} + \tilde \sigma_{\mu \nu}^{n_g; n_g+1, +}  \,.
\end{align}
We use the LLL self-energy from Eq.~\eqref{eqsiglll-nondeg} and the neighboring LL self-energy:
\begin{align}
\label{eqsig-delp}
\Sigma_{n_g+1, +}(\upmu) &  \approx
 L_x \int \frac{ dk_x'} {2 \, \pi}  \;
\frac{ \mathcal{M}_{k_x, k_x'}^{ n_g+1,+; n_g,+ } } { - \,\Sigma_{n_g, +} (\upmu)}  \,.
\end{align}
After combining with the vv selection rule, 
\begin{align}
\sigma_{\mu \nu}^{ql} =   \tilde \sigma_{\mu \nu}^{n_g; n_g+1, +}  \,.
\end{align}

\end{itemize}

\subsubsection{$2 \,E_g > \min\!\left\{ \Delta_+,\, \Delta_-\right\} $ and $ \rho = 2\, N + 1 $}
\label{sec-cond-gnr3}

Here, only positive-energy LLs (i.e., $ s = \tilde s = + $) are to be considered.
There is degeneracy of LLs indexed by ${n_g \pm N'}$ (except for the LLL) satisfying $ E_{n_g + N'} = E_{n_g - N'}$, as long as $n_g - N'\geq 0 $. For $n_g \geq 1 $: $\Delta_- = \Delta_+$. Spectral-weight overlaps with LLL to compute conductivity (3 terms in the sum for $\sigma_{xx}^{ql}$): $n_g \to n_g $, $n_g \to n_g-1$, and $n_g \to n_g+1$. The complete sets comprise $ \lbrace \psi_{n_g, +}, \,\psi_{n_g-1, +}, \,\psi_{n_g+1, +} \rbrace $ and $ \lbrace  E_{g} , \, E_{n_g-1} = E_{n_g+1}\rbrace$, leading to
\begin{align}
\sigma_{\mu \nu}^{ql} = \tilde \sigma_{\mu \nu}^{n_g ; n_g, +} + \tilde \sigma_{\mu \nu}^{n_g; n_g-1, +}  + \tilde \sigma_{\mu \nu}^{n_g; n_g+1, +}  \,.
\end{align}
This case is analogous to the Dirac case. We use Eqs.~\eqref{eqsiglll-nondeg}, \eqref{eqsig-delm}, and \eqref{eqsig-delp}.
After combining with the vv selection rule, 
\begin{align}
\sigma_{\mu \nu}^{ql} =  \tilde \sigma_{\mu \nu}^{n_g; n_g-1, +}  + \tilde \sigma_{\mu \nu}^{n_g; n_g+1, +}  \,.
\end{align}

\subsubsection{$2 \,E_g > \min\!\left\{ \Delta_+,\, \Delta_-\right\} $ and $ \rho = 2\, N  $}
\label{sec-cond-gnr4}

Here, only positive-energy LLs (i.e., $s = \tilde s = +$) are to be considered.
There is degeneracy of LLs indexed by $n_1 + N'$ and $n_1+1 - N'$ (including LLL), satisfying $ E_{n_1 + N'} = E_{n_1+1 - N'}$, as long as $n_1+1 - N' \geq 0 $.
For $n_1 \geq 1 $: $\Delta_- = \Delta_+$. Spectral-weight overlaps with LLL to compute conductivity (8 terms in the sum for $\sigma_{xx}^{ql}$): $n_1 \to n_1 $, $n_1 \to n_2 $, $n_1 \to n_1-1 $, $n_1 \to n_1+2 $, $n_2 \to n_2 $, $n_2 \to n_1 $, $n_2 \to n_1-1 $, $n_2 \to n_1+2 $.
The complete sets comprise $ \lbrace \psi_{n_1, +}, \,\psi_{n_1-1, +}, \,\psi_{n_2, +}, \,\psi_{n_2+1, +} \rbrace $ and $ \lbrace  E_{n_1} = E_{n_2} , \, E_{n_1-1} = E_{n_2 +1} \rbrace$, leading to
\begin{align}
\sigma_{\mu \nu}^{ql} = \tilde \sigma_{\mu \nu}^{n_1; n_1, +} + \tilde \sigma_{\mu \nu}^{n_1; n_2, +} +\tilde \sigma_{\mu \nu}^{n_2; n_1, +} + \tilde \sigma_{\mu \nu}^{n_2; n_2, +}
+ \tilde \sigma_{\mu \nu}^{n_1; n_1-1, +} + \tilde \sigma_{\mu \nu}^{n_1; n_2+1, +} +\tilde \sigma_{\mu \nu}^{n_2; n_1-1, +} 
+ \tilde \sigma_{\mu \nu}^{n_2; n_2+1, +} \,.
\end{align}



We use Eq.~\eqref{eqsiglll-deg} for the degenerate LLL self-energies and the following for the neighboring levels:
\begin{align}
\Sigma_{n_1-1, + }(\upmu) & \approx
 L_x \int \frac{ dk_x'} {2 \, \pi}  \;
\frac{ \mathcal{M}_{k_x, k_x'}^{n_1-1,+; n_1,+ } } {- \,\Sigma_{ n_1, +} (\upmu)}
+
L_x \int \frac{ dk_x'} {2 \, \pi}  \;
\frac{ \mathcal{M}_{k_x, k_x'}^{n_1 -1,+ ; n_1+1, +} } {  -\, \Sigma_{ n_1+1, +} (\upmu)}\,,
\nn \Sigma_{n_2+1, +}(\upmu) & \approx
 L_x \int \frac{ dk_x'} {2 \, \pi}  \;
\frac{ \mathcal{M}_{k_x, k_x'}^{n_1+2 , +; n_1 , +} } { - \,\Sigma_{ n_1, +} (\upmu)}
+
L_x \int \frac{ dk_x'} {2 \, \pi}  \;
\frac{ \mathcal{M}_{k_x, k_x'}^{n_1 +2, +; n_1+1, + } } {-\, \Sigma_{ n_1+1, +} (\upmu)}\,.
\end{align}
After combining with the vv selection rule, 
\begin{align}
\sigma_{\mu \nu}^{ql} =   \tilde \sigma_{\mu \nu}^{n_1; n_1+1, +} +\tilde \sigma_{\mu \nu}^{n_1+1; n_1, +} 
+ \tilde \sigma_{\mu \nu}^{n_1; n_1-1, +}  
+ \tilde \sigma_{\mu \nu}^{n_1+1; n_1+ 2, +} \,.
\end{align}

\section{Results and discussions}
\label{secres}

For obtaining representative plots depicting the nature of conductivity, we use realistic parameter values from the literature, as shown below.
\begin{table}[h!]
\centering
\begin{tabular}{|c|c|c|c|}
\hline
 Reference & Parameter & Traditional Units & Natural Units \\ \hline
 \cite{sras3-nodalring, yang_lieb_be2c_beh2} & $k_0$  & $0.07$--$0.76$ \AA$^{-1}$ & $134$--$1500$ eV\\ \hline
\cite{sras3-nodalring} & $m^*$ & $0.017$--$0.148\,m_e$ & $8.7\times 10^3$--$7.6\times 10^4$ eV\\ \hline
 \cite{jin_mx_nodeline,lu_kagome_honeycomb} & $\Delta$ & $0.005$--$0.034$ eV & $0.005$--$0.034$ eV\\ \hline
 \cite{dirac-qtm, adam-graphene} & $n_{\rm{imp}}$ & $2\times10^{-3}$ -- $4\times10^{-2}$ nm$^{-2}$ & $77.9$ -- $1.56 \times 10^3$ eV$^{2}$ \\ 
 & $V_0$ & $3.26$ eV nm$^{2}$ & $8.37 \times 10^{-5}$ eV$^{-1}$ \\
 & $n_{\rm{imp}}\, V_0^2$ & $0.0213$ -- $0.425$ eV$^{2}$ nm$^{2}$ & $5.46 \times 10^{-7}$ -- $1.09\times10^{-5}$ \\ 
 & $\epsilon$ & $6.9\,\epsilon_0$ & $6.9$ \\
 & $\tilde{V}_0$ & $3.44$ eV nm & $0.0175$ \\
 & $n_{\rm{imp}}\, {\tilde V_0}^2$ & $0.0237$ -- $0.474$ eV$^2$ & $0.0237$ -- $0.474$ eV$^2$ \\ \hline
\cite{minke-magnetotransport, dirac-qtm} &  $R$ & $0.44$--$8.5$ nm & $0.0022$--$0.0431$ eV$^{-1}$ \\ \hline 
\cite{adam-graphene, dassarma-graphene} & $\kappa$ & $0.1$--$0.5 $ nm$^{-1}$ & $19.73$--$98.65 $ eV\footnote{This range for $\kappa$ is a rough estimate from standard screening theory for graphene and typical carrier-densities discussed in Refs.~\cite{adam-graphene, dassarma-graphene}.} \\ \hline 
\end{tabular}
\caption{Realistic scales informing our choice of GNR parameters, converted to natural units (we have used $\hbar \,c = 1973$ eV\,\AA, $m_e\, c^2 = 5.11\times10^5$ eV).
\label{tab:params}}
\end{table}
We choose $k_0 = 200$ eV and $m^* = 2 \times 10^4 $ eV, which sets the energy scale $\varepsilon_r = 1$ eV. The gap is scanned over the narrow band $\Delta=0.0335$--$0.034$ eV used throughout the figures below. The plots scan the dimensionless variable $\rho=2\varepsilon_r/\varepsilon_c$ over the window $\rho\in[20,50]$. This follows directly from the structure of $E_g$ in Eq.~\eqref{eq:gnr-spectrum}. Near a level crossing, $(2n+1-\rho)\to0$ and $E_g\to\Delta$, and the LLL energy scale that controls the resonance condition therefore inherits the smallness of the row-to-row spread in $\Delta$ itself. Since this spread is far below the disorder-induced linewidths $\Gamma_n$ explored below, the oscillatory structure of both $\sigma_{xx}^{ql}$ and $\sigma_{xy}^{ql}$ is essentially set by $\rho$ and by the disorder parameters alone, and is insensitive to $\Delta$ over this range.

\subsection{Longitudinal conductivity}
\label{secsxx}

\begin{figure}[t!]
\centering
\includegraphics[width= 0.85\linewidth]{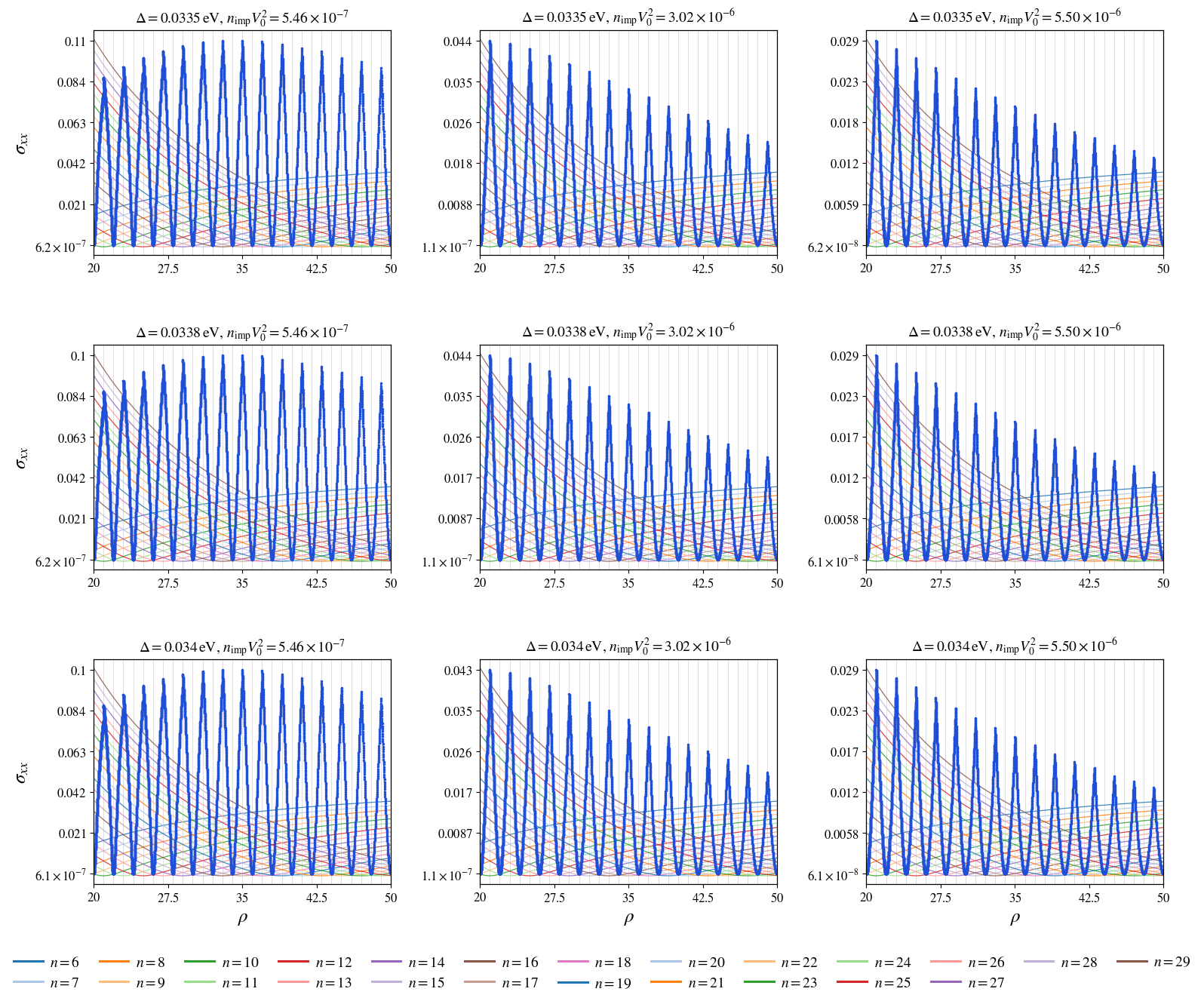}
\caption{GNR: $\sigma_{xx}^{ql}$ versus $\rho$ for pointlike impurities. The rows correspond to different values of the gap, $\Delta $, while the columns contain varying values of the effective disorder strength, $n_{\mathrm{imp}} \,V_0^2 $. The coloured curves in the background trace the evolution of LLs with $\rho$, for $n$ ranging from 6 to 29.\label{figsxx1}}
\end{figure}

\begin{figure}[ht!]
 \centering
\includegraphics[width= 0.85\linewidth]{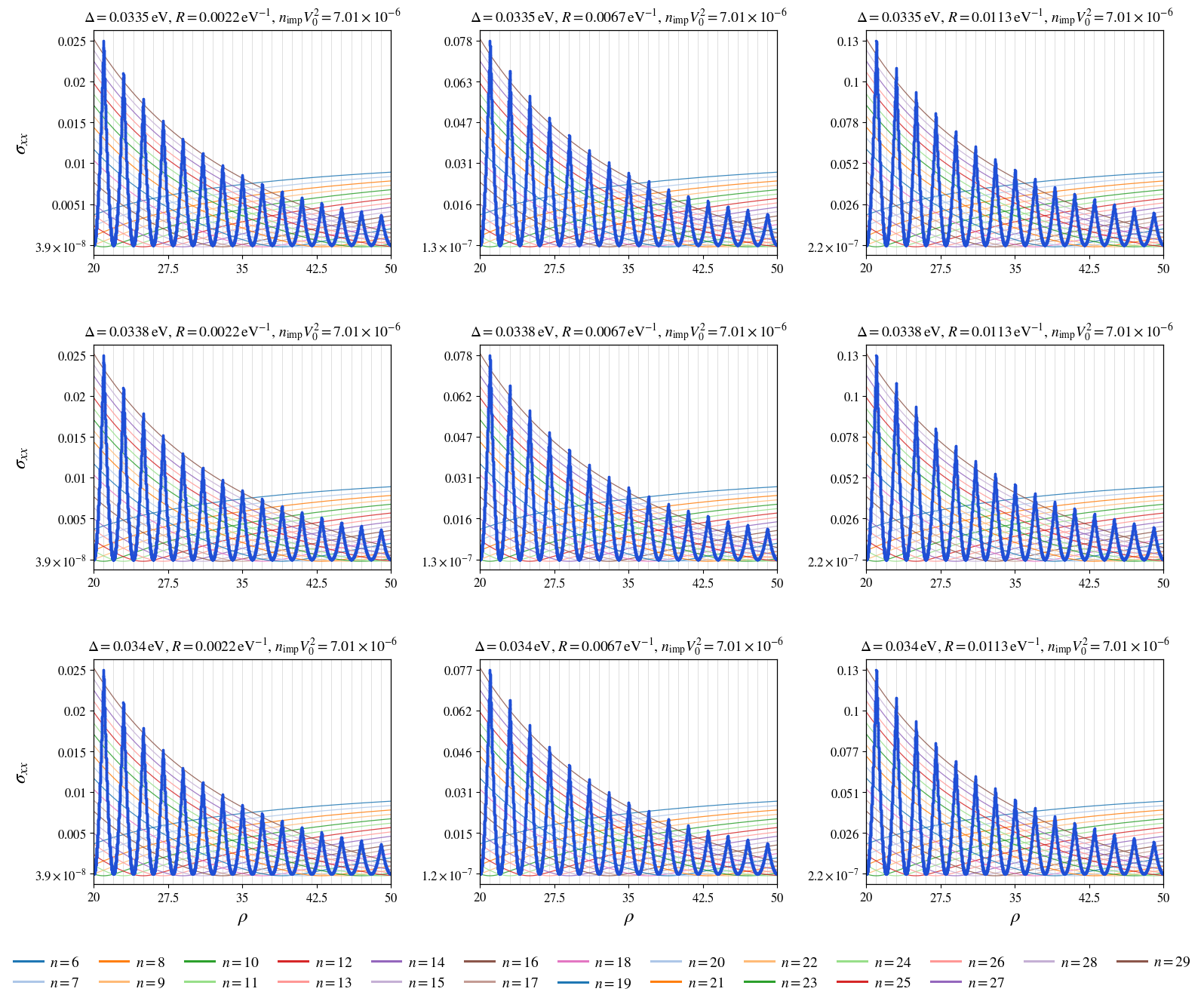}
\caption{GNR: $\sigma_{xx}^{ql}$ versus $\rho$ for Gaussian impurity-potentials. The rows correspond to different values of the gap, $\Delta $, while the columns contain varying values of $ R $. The coloured curves in the background trace the evolution of LLs with $\rho$, for $n$ ranging from 6 to 29.\label{figsxx2}}
\end{figure}

Figs.~\ref{figsxx1}, \ref{figsxx2}, and \ref{figsxx3} show the ultraquantum-limit longitudinal conductivity $\sigma_{xx}^{ql}(\rho)$ [cf. Eq.~\eqref{eq:sigxx-ql}], evaluated using Eqs.~\eqref{eqsel_LLL} and \eqref{eqsel_neigh}, for the three impurity models. The coloured background lines in each panel trace a selection of the LLs as functions of $\rho$, the grey vertical lines mark the integer values of $\rho$ at which a level becomes doubly degenerate, and the blue curve is $\sigma_{xx}^{ql}(\rho)$ itself. A useful check across the three figures is that, for a given $\Delta$, the peaks of $\sigma_{xx}^{ql}$ sit at exactly the same values of $\rho$ regardless of which impurity model is used. This is expected, since the crossing condition depends only on the LL spectrum and not on the disorder, and it confirms that what the three models change is purely the height and width of the resonances rather than where they occur.

As $B$ grows, $\rho$ falls, and every time it crosses one of the grey lines the level currently playing the role of the LLL becomes momentarily degenerate with its neighbour before handing the role over to it. It is precisely at these crossings that $\mathcal A_n$ peaks for both the LLL and its neighbour, which is why $\sigma_{xx}^{ql}$ develops a resonant spike each time. Right at degeneracy, since the peak height scales roughly as $1/(\Gamma_{n_g}\Gamma_n)$, narrower disorder-induced linewidths give taller and sharper resonances. This is why the same comb of peaks appears far more pronounced in the cleaner columns of Figs.~\ref{figsxx1}--\ref{figsxx3} than in the more strongly-disordered ones.

Within the window $\rho\in[20,50]$ used in the plots, the doubly-degenerate points are located at $\rho=2N$ for integer $N$ and are therefore equally spaced in $\rho$, with spacing $\Delta\rho=2$; this is exactly what the grey vertical lines show. What varies from peak to peak, and from panel to panel, is not this spacing but the height and width of the resonance sitting at each crossing, both of which are controlled by $\Gamma_{n_g}(\rho)$ and $\Gamma_{n}(\rho)$, quantities that depend on $\rho$ and on $n$ through the disorder-averaged self-energy. Owing to this $\rho$- and $n$-dependence, some crossings produce tall, narrow spikes while others produce comparatively broad or subdued ones, and the resulting pattern differs from column to column, as can be seen by comparing the panels within Figs.~\ref{figsxx1}, \ref{figsxx2}, and \ref{figsxx3}.

\subsubsection{Short-ranged potential}

For pointlike impurities, represented by Fig.~\ref{figsxx1}, solving the self-consistent Eq.~\eqref{eqsel_LLL} for the LLL gives $\Gamma_{n_g}\simeq\sqrt{n_{\rm imp}V_0^2/(2\pi\ell_B^2)}$. Clearly, this result carries no dependence on $n_g$ itself. It follows from the momentum-independence of the white-noise correlator: once the $k_x'$-integral in Eq.~\eqref{eqselfen} is carried out, the value, $\mathcal C \equiv n_{\rm imp}V_0^2/(2\pi\ell_B^2)$ no longer retains any trace of the LL index it was computed for. The same value also sets the linewidth of the neighbouring level, although the two broadenings come from different equations --- the LLL self-energy follows from the self-consistent Eq.~\eqref{eqsel_LLL}, while the neighbour's self-energy follows from the non-self-consistent Eq.~\eqref{eqsel_neigh}, evaluated using the already-determined $\Sigma_{n_g}$. For pointlike impurities, the matrix-element entering both equations reduces to $\mathcal C$, giving $\Sigma_{n_{gn}}=i\,\Gamma_{n_g}$ and therefore $\Gamma_{n_{gn}}=\Gamma_{n_g}$ exactly, for white-noise disorder. The peak height should then scale as $1/\Gamma_{n_g}^2\propto1/(n_{\rm imp}V_0^2)$.

The three columns of Fig.~\ref{figsxx1} give a tallest peak of $\sigma_{xx}^{ql}\approx0.11$ at $n_{\rm imp}V_0^2=5.46\times10^{-7}$, falling to $\approx0.044$ at $3.02\times10^{-6}$ and to $\approx0.029$ at $5.50\times10^{-6}$, successive ratios of about $2.5$ and $1.5$. These fall noticeably short of the ratios of the coupling constants themselves, about $5.5$ and $1.8$. Part of the shortfall traces back to where each tallest peak sits: for the two more disordered columns it occurs close to the left edge of the window, near $\rho\approx20.5$, whereas for the weakest-disorder column it occurs further into the window, closer to $\rho\approx30$. Because the peak amplitude also carries the $B^2\max(n_0,n)$ prefactor of $\tilde\sigma_{xx}^{n_0;n,s}(\upmu)$ in Eq.~\eqref{eq:sigxx-ql} (via the matrix-element of Eq.~\eqref{eqdotsq} with $\tilde n\to n_0$), comparing peaks located at different $\rho$ compares different prefactors along with different $\Gamma_{n_g}$, which is why the observed ratios undershoot the pure $1/\Gamma_{n_g}^2$ expectation. For the two more-disordered columns, whose tallest peaks both sit near $\rho\approx20.5$, the predicted ratio of $1.8$ sits reasonably close to the observed $1.5$.

The prefactor $B^2\max(n_0,n)$, entering Eq.~\eqref{eq:sigxx-ql} through Eq.~\eqref{eqdotsq}, does not push the peaks to grow monotonically across the whole window. Its explicit $B^2$ dependence favours taller peaks as $\rho$ falls, an effect partly offset by $\max(n_0,n)$, which itself falls as $B$ increases since $n_g$ steps down towards zero over this range, and further offset by the growth of $\Gamma_{n_g}$ with $\sqrt{B}$. For the weakest-disorder column, this balance produces an envelope that rises from the $\rho=20$ edge to a maximum around the middle of the window before declining gently towards $\rho=50$. For the two more disordered columns, where $\Gamma_{n_g}$ starts out larger, its growth already dominates from the outset, the tallest peak sits right at the $\rho=20$ edge, and peak-heights fall off monotonically as $\rho$ increases. Away from a crossing, $\upmu\simeq E_g$ throughout the ultraquantum limit, so the LLL's own factor $\mathcal A_{n_g,+}(\upmu)$ stays essentially pinned at its resonant value $2/\Gamma_{n_g}$, while the neighbour's factor sees a detuning $(\upmu-E_n)^2\simeq\Delta_\pm^2$. Provided $\Delta_\pm\gg\Gamma_{n_g}$, the $\Gamma_{n_g}^2$ term in the denominator of $\mathcal A_{n,+}(\upmu)$ is negligible, leaving $\mathcal A_{n,+}(\upmu)\approx2\,\Gamma_{n_g}/\Delta_\pm^2$ (where we have used $\Gamma_{n_{gn}}=\Gamma_{n_g}$, as applicable for pointlike disorder). The trough of $\sigma_{xx}^{ql}$ is then set by the product of these two factors, $\mathcal A_{n_g,+}\; \mathcal A_{n,+}\sim4/\Delta_\pm^2$, which stays comparatively flat and close to zero on the linear axis throughout, since it carries no divergence as $\rho$ moves away from a crossing.

\subsubsection{Long-ranged potentials}

\begin{figure}[t!]
    \centering
    \includegraphics[width= 0.85\linewidth]{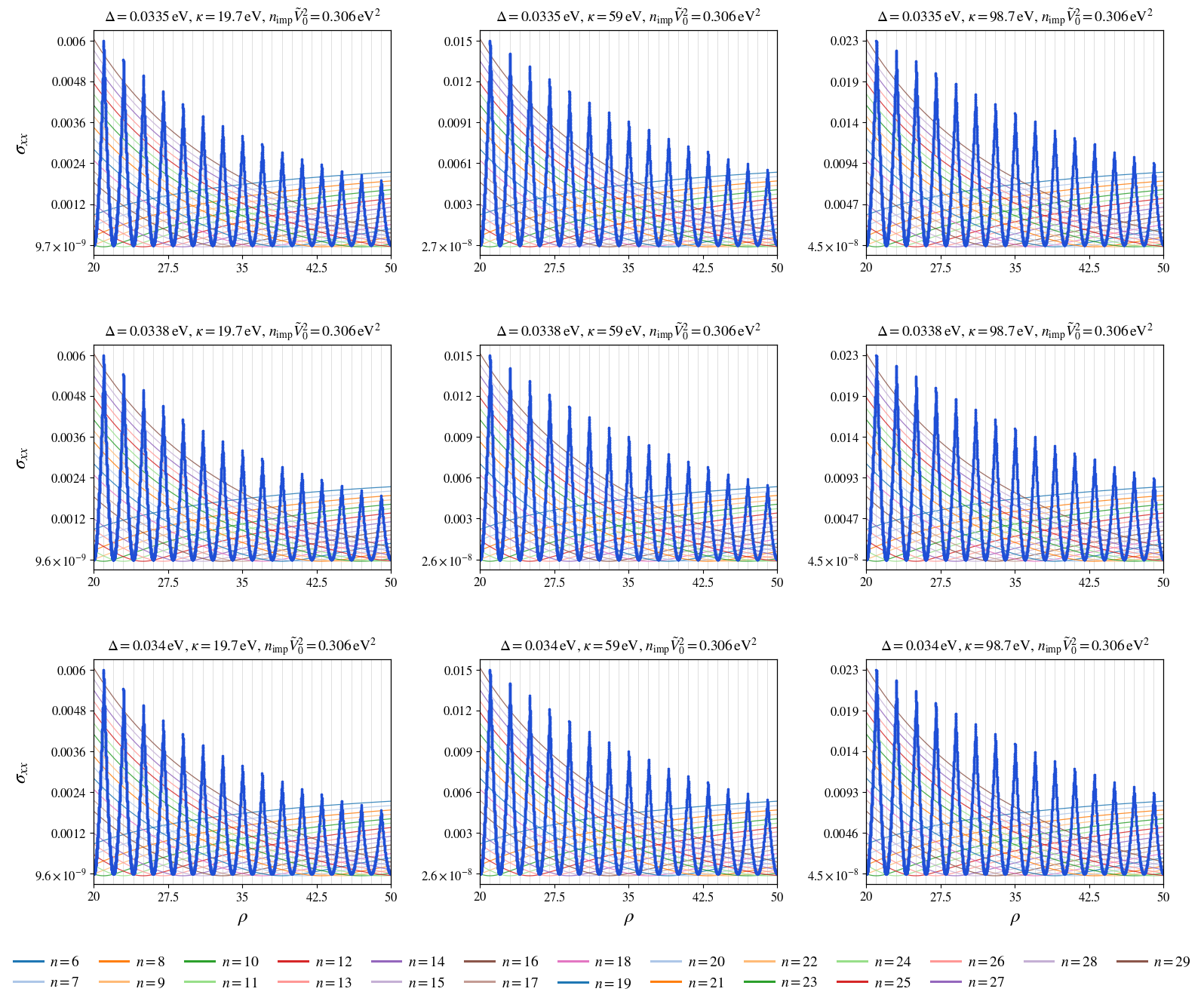}
\caption{GNR: $\sigma_{xx}^{ql}$ versus $\rho$ for Yukawa-type impurity-potentials. The rows correspond to different values of the gap, $\Delta $, while the columns contain varying values of $ \kappa $. The coloured curves in the background trace the evolution of LLs with $\rho$, for $n$ ranging from 6 to 29.\label{figsxx3}}
\end{figure}

The Gaussian and Yukawa panels of Figs.~\ref{figsxx2} and \ref{figsxx3} look qualitatively different from the pointlike case. A long-ranged potential carries an explicit LL-dependence in its matrix-element, built from Jacobi polynomials for the Gaussian potential [cf. Eq.~\eqref{eq:self-energy-final}] and from Laguerre polynomials for the Yukawa potential [cf. Eqs.~\eqref{eq:yukawa-nn}--\eqref{eq:yukawa-adj}], and this breaks the equality $\Gamma_{n_{gn}}=\Gamma_{n_g}$ found for white-noise disorder. Over the parameter range considered here, $\Gamma_n$ falls steadily as $n$ increases. A high-index orbital has a larger guiding-centre radius $\sim \ell_B \, \sqrt{n}$ and it, therefore, samples a smooth potential over a wider area, which weakens the scattering it feels. This $n$-dependence reshapes the envelope of $\sigma_{xx}^{ql}$ completely. At large $\rho$ the level $n_g$ is large, $\Gamma_{n_g}$ is small, and $\sigma_{xx}^{ql}$ stays correspondingly small. As $\rho$ falls and $n_g$ moves towards zero, $\Gamma_{n_g}$ grows and the conductivity climbs steeply across the whole window, far more steeply than for pointlike disorder (where $\Gamma_{n_g}$ carried no $n_g$-dependence and the rise was correspondingly gentle). This is why the upper envelope in Figs.~\ref{figsxx2} and \ref{figsxx3} spans several orders of magnitude, against the single order of magnitude spanned in Fig.~\ref{figsxx1}.

The parameters $R$ and $\kappa$ set the range of the two potentials, and through it the scale of the resonances. A larger $R$ lengthens the Gaussian potential, while a larger $\kappa$ shortens the Yukawa potential. In each case, the potential transfers less momentum on average, backscattering into the same orbital weakens, and $\Gamma_{n_g}$ shrinks. Because the resonant peak height scales roughly as $1/\Gamma_{n_g}^2$, the resonances grow taller and sharper as either parameter increases. Fig.~\ref{figsxx2} confirms this trend: the tallest peak rises from $\sigma_{xx}^{ql}\approx0.025$ at $R=0.0022\ \text{eV}^{-1}$ to $\approx0.078$ at $R=0.0067\ \text{eV}^{-1}$ and to $\approx0.13$ at $R=0.0113\ \text{eV}^{-1}$, an overall factor of about five. Fig.~\ref{figsxx3} shows the same pattern for the Yukawa potential, with the tallest peak rising from $\approx0.006$ at $\kappa=19.7$ eV to $\approx0.015$ at $\kappa=59$ eV and to $\approx0.023$ at $\kappa=98.7$ eV, a factor of about 3.8. Both increases are sublinear in the tuning parameter, because the polynomial prefactors of Eq.~\eqref{eq:self-energy-final} and Eqs.~\eqref{eq:yukawa-nn}--\eqref{eq:yukawa-adj} depend weakly on $R$ and $\kappa$ through the combinations $S=2R^2/\ell_B^2+1$ and $\beta=\kappa^2 \, \ell_B^2/2$, which partially offset the pure $\Gamma_{n_g}^{-2}$ scaling.

Although $n_{\rm imp}$ is held fixed in Figs.~\ref{figsxx2} and \ref{figsxx3}, with only $R$ and $\kappa$ scanned there, its role can be inferred from the pointlike results of Fig.~\ref{figsxx1}. The Gaussian and Yukawa matrix-elements carry a strong $n$-dependence through their polynomials, yet the self-consistent solution of Eqs.~\eqref{eqsel_LLL} and \eqref{eqsel_neigh} still gives a linewidth scaling as $\Gamma_n\propto\sqrt{n_{\rm imp}}$ at fixed $n$, exactly as for pointlike disorder, with only the $n$-dependent prefactor differing. The peak-height scaling should therefore follow the same $\sim1/n_{\rm imp}$ trend found for pointlike disorder, though that trend was itself only approximate there, since the prefactor effects discussed above already made the observed pointlike ratios fall short of the naive expectation. Increasing $n_{\rm imp}$ is expected to rescale the whole curve downward while leaving the resonance positions, fixed purely by the level-crossing condition $\Delta_\pm\to0$, unaffected.

\subsubsection{Comparison across disorder types}
\label{sec:res-comparison}

The three impurity models share the same resonance mechanism and differ only in how that mechanism is dressed. Every peak in Figs.~\ref{figsxx1}--\ref{figsxx3} owes its height to the same product $\Gamma_{n_g}\,\Gamma_{n_{gn}}$, and what separates pointlike scattering from the two long-ranged potentials is purely the $n$-dependence carried by this product. Pointlike disorder leaves no trace of $n$ in $\Gamma_n$, collapsing to $\Gamma_{n_g}^2$, and produces a comb whose amplitude is fixed by $n_{\rm imp}V_0^2$ alone and whose envelope climbs by roughly a factor of ten across the field window. The Gaussian and Yukawa potentials instead sample the guiding-centre wavefunction unevenly across $n$: a high-index orbit has a wide guiding-centre radius, senses a smooth potential only weakly, and carries a correspondingly small $\Gamma_n$. Since $n_g$ starts large at weak field and falls towards zero as $B$ grows, this $n$-dependence pins the conductivity down for most of the window and lets it climb sharply only once $n_g$ has fallen close to zero, giving the much steeper rise, spanning several orders of magnitude, seen in Figs.~\ref{figsxx2} and \ref{figsxx3}.
 
Within either long-ranged model, the range parameter sets the overall scale of the envelope rather than its shape. A larger $R$ or $\kappa$ weakens the momentum a scattering event can transfer, narrows $\Gamma_{n_g}$, and sharpens every resonance in the comb, as demonstrated quantitatively above. The impurity density plays a cleaner and disorder-independent role: because $\Gamma_n\propto\sqrt{n_{\rm imp}}$ at every Landau-level index regardless of the potential shape, raising $n_{\rm imp}$ rescales the whole curve down by roughly $1/n_{\rm imp}$ and leaves both the resonance positions and the shape of the envelope untouched, since those positions are fixed purely by the level-crossing condition and carry no dependence on disorder at all. What survives across all three models, then, is a clean separation between a fine comb that is a disorder-independent fingerprint of the field-driven migration of $n_g(B)$ through the checkmark spectrum, and a slowly varying envelope whose shape is set by the range of the impurity potential and whose amplitude is set by the effective disorder strength, $n_{\rm imp}V_0^2$ for the pointlike and Gaussian cases and $n_{\rm imp}\tilde V_0^2$ for the Yukawa potential.

\subsection{Hall conductivity}
\label{secxy}

\begin{figure}[t!]
    \centering
    \includegraphics[width=0.85 \linewidth]{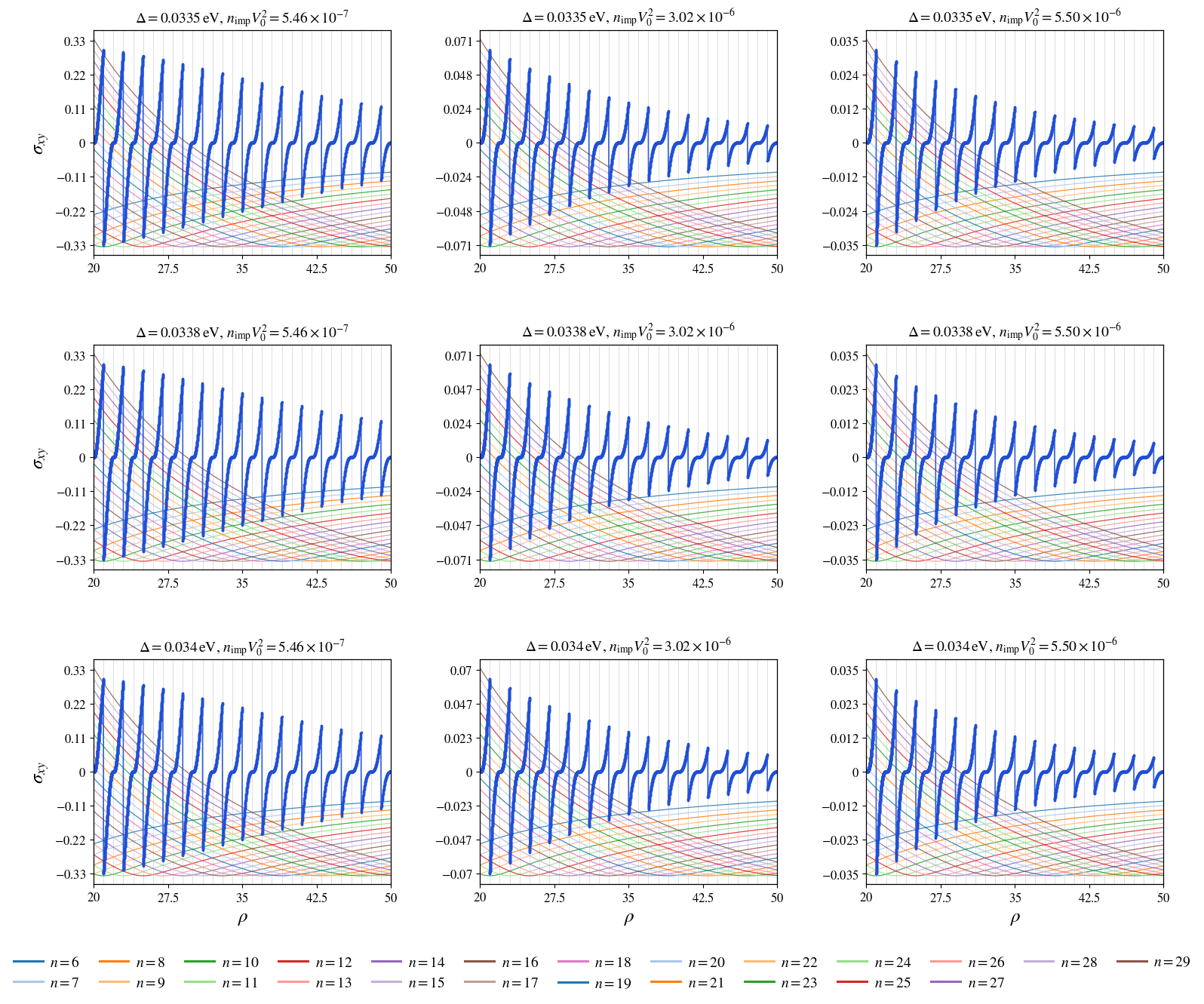}
\caption{GNR: $\sigma_{xy}^{ql}$ versus $\rho$ for pointlike impurities. The rows correspond to different values of the gap, $\Delta $, while the columns contain varying values of the effective disorder strength, $n_{\mathrm{imp}} \,V_0^2 $. The coloured curves in the background trace the evolution of LLs with $\rho$, for $n$ ranging from 6 to 29. Crucially, the sharp zero crossings of the sawtooth patterns align perfectly with the grey vertical lines representing $\rho = 2N$. The alternating signs of the teeth arise from the $\mathrm{sgn}(n-n_0)$ factor in Eq.~\eqref{eq:sigxy-ql-final}.
\label{figsxy1}}
\end{figure}

\begin{figure}[t!]
    \centering
    \includegraphics[width=0.85 \linewidth]{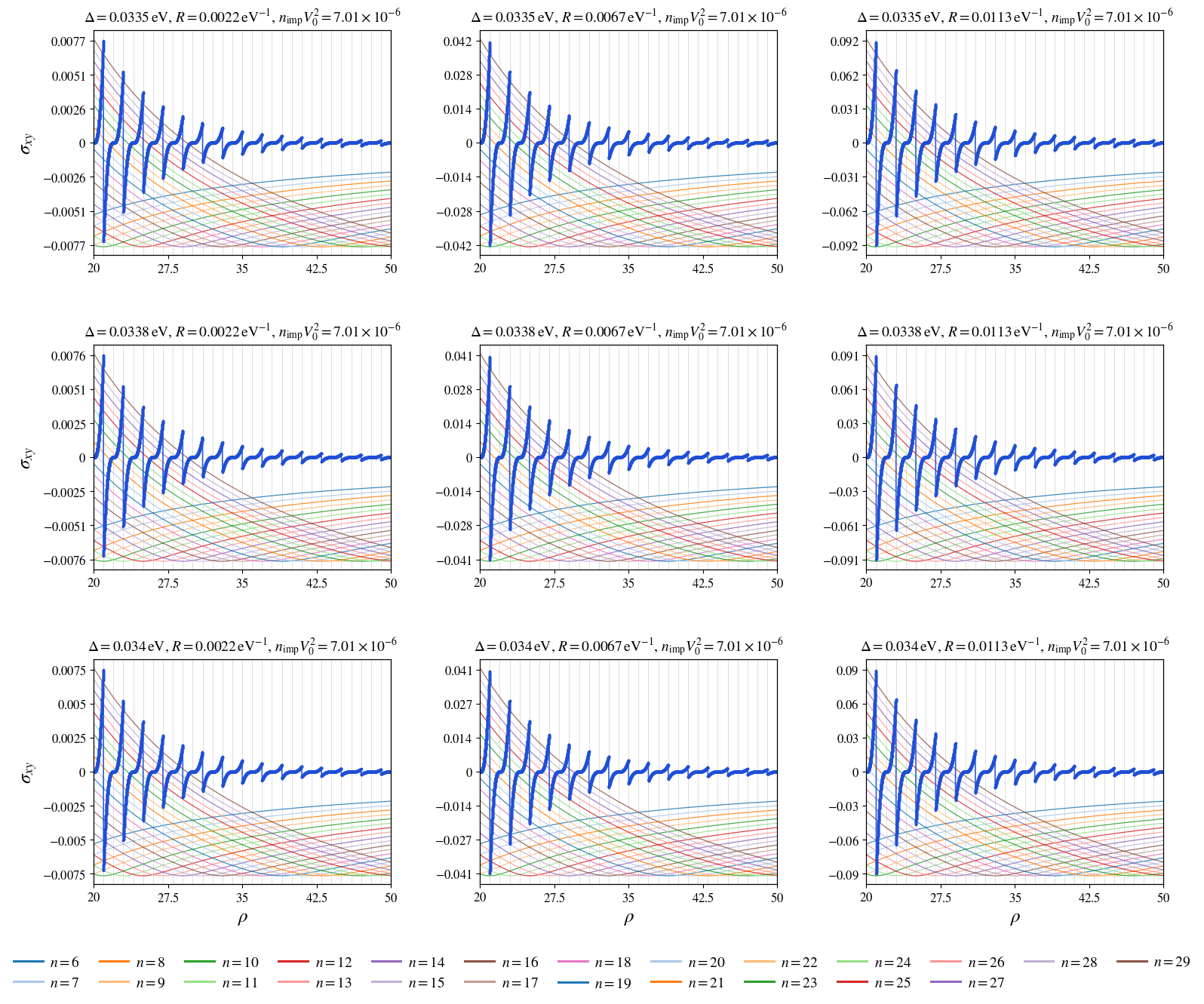}
\caption{GNR: $\sigma_{xy}^{ql}$ versus $\rho$ for Gaussian impurity potentials. The rows correspond to different values of the gap, $\Delta $, while the columns contain varying values of $ R $. The coloured curves in the background trace the evolution of LLs with $\rho$, for $n$ ranging from 6 to 29. The sharp zero crossings of the sawtooth pattern align with the grey vertical lines representing $\rho = 2N$, and the alternating sign of the teeth arises from the $\mathrm{sgn}(n-n_0)$ factor in Eq.~\eqref{eq:sigxy-ql-final}.
\label{figsxy2}}
\end{figure}

\begin{figure}[t!]
    \centering
    \includegraphics[width=0.85 \linewidth]{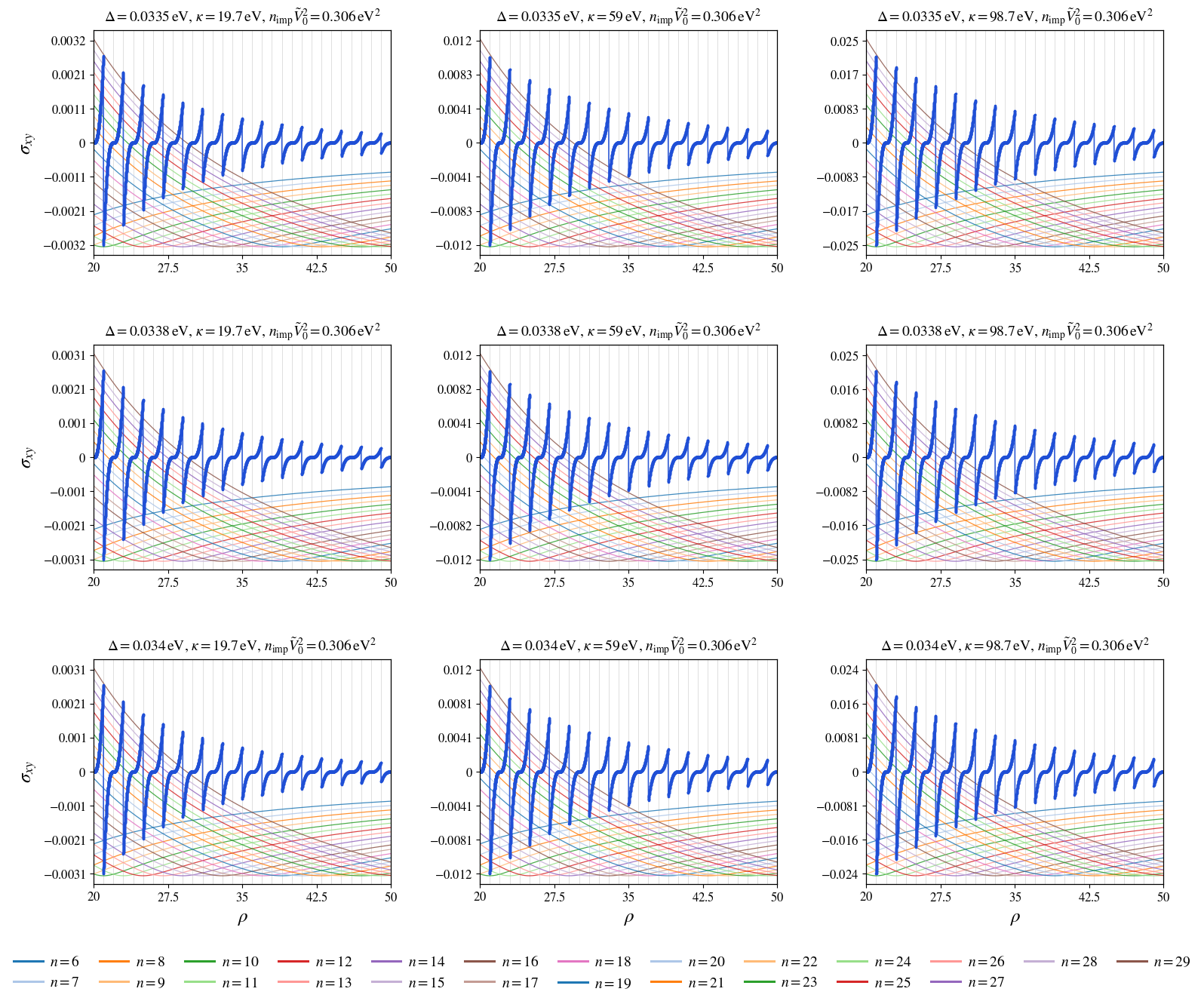}
\caption{GNR: $\sigma_{xy}^{ql}$ versus $\rho$ for Yukawa-type impurity potentials. The rows correspond to different values of the gap, $\Delta $, while the columns contain varying values of $ \kappa $. The coloured curves in the background trace the evolution of LLs with $\rho$, for $n$ ranging from 6 to 29. The sharp zero crossings of the sawtooth pattern align with the grey vertical lines representing $\rho = 2N$, and the alternating sign of the teeth arises from the $\mathrm{sgn}(n-n_0)$ factor in Eq.~\eqref{eq:sigxy-ql-final}.
\label{figsxy3}}
\end{figure}

Figs.~\ref{figsxy1}, \ref{figsxy2}, and \ref{figsxy3} show $\sigma_{xy}^{ql}(\rho)$ for the same three impurity models, computed from Eq.~\eqref{eq:sigxy-ql-final}. Where $\sigma_{xx}^{ql}(\rho)$ is a comb of positive resonant spikes, $\sigma_{xy}^{ql}$ traces out a sawtooth about zero, with its zero-crossings pinned exactly to the grey guiding lines at $\rho=2N$. The reason is contained in Eq.~\eqref{eq:sigxy-ql-final} itself. Away from a crossing, only one neighbouring level feeds the sum, whichever of $n_g-1$ or $n_g+1$ sits closer in energy, and the factor $\mathrm{sgn}(n-n_0)$ fixes the sign of the resulting tooth according to which side of $n_g$ that neighbour sits on. Each time $\rho$ passes through an even integer, the two neighbours become momentarily degenerate (i.e., $\Delta_-=\Delta_+=0$), their contributions enter the sum with equal weight but opposite signs, and $\sigma_{xy}^{ql}$ is carried smoothly through zero before the dominant neighbour switches sides and the next tooth begins. This is a genuinely different mechanism from the one governing $\sigma_{xx}^{ql}$, where both the spectral factors entering the product are resonant Lorentzians of fixed sign. On the contrary, for $\sigma_{xy}^{ql}$, only $\mathcal A_{n_0,+}$ is resonant, while $\mathcal B_{n,s}$ is dispersive and switches sign as $\upmu$ sweeps past $E_n$, which is what makes the curve change sign from one tooth to the next.

For pointlike impurities, the teeth grow shorter with increasing $\rho$, tracking the diminishing envelope of $\sigma_{xx}^{ql}$. This is dictated by the growth of $\Gamma_{n_g}$ with $\sqrt B$, together with the $B^2\max(n_0,n)$ prefactor of Eq.~\eqref{eqdotsq}. Raising the disorder strength broadens each resonance and suppresses the tooth-heights, from a tallest tooth of $\sigma_{xy}^{ql}\approx0.33$ at $n_{\rm imp}V_0^2=5.46\times10^{-7}$ down to $\approx0.071$ at $3.02\times10^{-6}$ and to $\approx0.035$ at $5.50\times10^{-6}$. This fall is a little steeper than the $1/\Gamma_{n_g}^2$ scaling found for $\sigma_{xx}^{ql}$, because the Hall response inherits a dispersive $\mathcal B_{n,s}$ factor in place of the second resonant $\mathcal A_{n,s}$ that appears in the longitudinal case.

The Gaussian and Yukawa teeth are small and tightly-spaced at large $\rho$, where $n_g$ is large and $\Gamma_{n_g}$ small. They grow taller and more widely-spaced as $\rho$ falls and $n_g$ drops with it, the same $n$-dependence of $\Gamma_n$ responsible for the steep rise of $\sigma_{xx}^{ql}$ towards the low-$\rho$ end of the window. The range parameters set the same overall scale. The tallest tooth in Fig.~\ref{figsxy2} grows from $\approx0.0077$ at $R=0.0022\text{ eV}^{-1}$ to $\approx0.042$ at $R=0.0067\text{ eV}^{-1}$ and to $\approx0.092$ at $R=0.0113\text{ eV}^{-1}$. In Fig.~\ref{figsxy3}, it grows from $\approx0.0032$ at $\kappa=19.7$ eV to $\approx0.012$ at $\kappa=59$ eV and to $\approx0.025$ at $\kappa=98.7$ eV, tracking the growth of the corresponding $\sigma_{xx}^{ql}$ peaks.

The Hall response is therefore a second fingerprint of the same physics as the longitudinal one. Both trace the field-driven migration of $n_g(B)$ through the checkmark spectrum. While $\sigma_{xx}^{ql}$ reports each crossing as a resonant spike, $\sigma_{xy}^{ql}$ reports it as a change of sign, with the zero-crossing itself marking the instant the LLL hands off its role to its neighbour. As with the longitudinal response, a larger $n_{\rm imp}$ is expected to rescale the entire sawtooth downward while leaving the zero-crossings pinned to $\rho=2N$, since those crossings are fixed purely by the level-crossing condition and not by disorder, though $n_{\rm imp}$ itself is held fixed throughout Figs.~\ref{figsxy2} and \ref{figsxy3} and this scaling is not tested directly here.
 
\subsection{Comparison with the Dirac case}
\label{sec:dirac-contrast}
 
The Dirac cone offers the clean counterpoint to all of this, because its LLL sits at the zeroth index for every value of the field. The details are provided in Appendix~\ref{appdirac}.
Unlike the LLL, whose self-energy $\Sigma_0$ is determined self-consistently because the LLL itself constitutes the dominant intermediate state [cf.\ Eq.~\eqref{eqsel_LLL} for the GNR], the self-energy $\Sigma_1$ of the neighbouring level must be evaluated non-self-consistently, using the already determined $G_0$ rather than $G_1$ [cf.\ Eq.~\eqref{eqsel_neigh}]. Thus, $\Sigma_1(\upmu)= {\vartheta_1}/{\upmu-\Sigma_0(\upmu)}$, and not $\Sigma_1=\vartheta_1/(\upmu-\Sigma_1)$, as was erroneously implemented in Ref.~\cite{dirac-qtm}. This is the main reason that we redo the Dirac-node calculations here. The qualitative features predicted in Ref.~\cite{dirac-qtm} remains unchanged, but the numerical prefactors change after out corrections.

The pair of neighbouring levels that dominates the ultraquantum limit formthe Dirac node never changes. Indeed, the resistivity is a smooth monotonic function of $B$ whose form is dictated entirely by the disorder type, given in Eq.~\eqref{eq:rho-scaling}. It stays field-independent for pointlike disorder, rises as $B^{1/2}$ for a Gaussian potential once its range exceeds the magnetic length and flattens towards a constant with a linear-in-$B$ correction in the opposite limit, and follows the same two regimes for the Yukawa potential depending on whether the screening length is held fixed or is instead determined self-consistently through RPA, as in Ref.~\cite{dirac-qtm}. The Hall conductivity, meanwhile, vanishes identically for all three disorder models, a direct consequence of the particle-hole symmetry relating the two LLs that flank the zeroth one. This is where the two systems part company most sharply. The Dirac cone gives smooth monotonic longitudinal transport and an identically vanishing Hall response, and it does so for any of the three impurity potentials and at any disorder strength or range. The GNR instead oscillates in both channels, resonant in $\sigma_{xx}^{ql}$ and sign-alternating in $\sigma_{xy}^{ql}$, at exactly the same fields where the LLL becomes degenerate with a neighbour. Since both features trace back to the field-driven migration of the LLL through the checkmark spectrum rather than to any particular choice of disorder strength, range, or gap, they persist unchanged across every parameter combination explored in Figs.~\ref{figsxx1}--\ref{figsxy3}. The oscillatory fingerprint of the GNR is therefore a consequence of its LL structure, rather than an artefact of the parameters chosen to illustrate it.
 
\section{Summary and concluding remarks}
\label{seccon}
 
We have studied the ultraquantum-limit dc magnetoconductivity of 2D Dirac cones and GNRs, where only the LLL is partially occupied. Working within the Kubo--Bastin formalism and a self-consistent Born treatment of disorder for the LLL, we obtained the longitudinal and Hall conductivities for pointlike, Gaussian, and Yukawa impurity potentials. The Dirac cone offers the simpler case: its LLL sits fixed at $n=0$ for every field, the resistivity is a smooth monotonic function of $B$ whose form depends on the disorder type alone, and the Hall conductivity vanishes for all three potentials as a direct consequence of the particle-hole symmetry relating the two levels adjacent to the LLL. The GNR behaves quite differently, because its Landau spectrum is non-monotonic in the level index. As the field grows, the level playing the role of the LLL steps down to progressively lower indices, and each time two levels become momentarily degenerate the longitudinal conductivity develops a resonant spike while the Hall conductivity passes through zero. Away from these degeneracies $\sigma_{xy}^{ql}$ traces out a sawtooth whose sign is fixed by whichever neighbouring level dominates. This oscillatory structure is a robust feature of the GNR, present for all three disorder models and traceable directly to the field-driven migration of the LLL, with disorder type and density affecting only the envelope that rides on top of it. Pointlike disorder gives a level-independent linewidth and a gently rising envelope, whereas the Gaussian and Yukawa potentials give a linewidth that narrows with increasing level index, producing weak conductivity at low field followed by a much steeper rise as the LLL index falls. A larger impurity density broadens the resonances and suppresses their height in every case, without shifting where they occur, and within the narrow range of gap values considered here changing $\Delta$ has little qualitative bearing on either the resonance positions or the sawtooth structure.
 
Taken together, these results offer a fairly clean diagnostic. An unstructured, monotonic longitudinal conductivity paired with a vanishing Hall response points to an ordinary Dirac-like node, while an oscillatory longitudinal conductivity paired with a sign-alternating Hall sawtooth is the signature of a GNR. Candidate materials for the latter include Kagome-honeycomb lattices~\cite{lu_kagome_honeycomb}, Be$_2$C and BeH$_2$ monolayers~\cite{yang_lieb_be2c_beh2}, and the MX family of compounds~\cite{jin_mx_nodeline}, and the effect should be most pronounced deep in the quantum limit, at the highest attainable fields.
 
A caveat is worth stating plainly. The $\sigma_{xy}^{ql}$ computed here is only part of the full dc Hall response. In the Smr\v{c}ka--\v{S}treda decomposition~\cite{smrcka, bonbien_manchon}, $\sigma_{xy}=\sigma_{xy}^{\rm I}+\sigma_{xy}^{\rm II}$, and our Kubo--Bastin trace of Eq.~\eqref{eq:KB} is built from the resonant $G^RG^A$ channel weighted by the thermal window $-\partial_E f$, which gives exclusively $\sigma_{xy}^{\rm I}$, the disorder-broadened Fermi-surface response that vanishes identically for the Dirac cone and produces the sign-alternating sawtooth for the GNR. Left aside throughout is $\sigma_{xy}^{\rm II}$, a $G^R$-only object weighted by the full Fermi function rather than by $-\partial_E f$, which survives even as the disorder broadening vanishes and is the Landau-level avatar of the Berry-curvature sum over the entire filled Fermi sea~\cite{tknn_iqh}. Taken at face value this sum diverges, because the hole branch is filled at every level index, but the divergence is spurious and cancels once the sum is regularised against its value at charge neutrality, leaving a finite, quantised count through the \v{S}treda formula $\sigma_{xy}^{\rm II}=e\,\partial n/\partial B|_\upmu$. For the GNR, this clean-limit Fermi-sea piece, its Chern number set by two field-dependent turning indices on the checkmark spectrum and the bumpy plateaus it produces as $B$ is swept at fixed $\upmu$, is exactly the quantity already worked out in Ref.~\cite{barati-nlsm-qhe}, and Appendix~\ref{app:sigmaII} recasts that result in our regularised-sum language. It rides on top of, and is essentially independent of, the disorder-broadened sawtooth $\sigma_{xy}^{\rm I}$ that is the subject of this paper: theirs is the clean-limit, topological Fermi-sea response, ours is the disorder-broadened Fermi-surface response, and together the two give the full $\sigma_{xy}$.
 
A few extensions follow naturally from this work. The calculation rests on a single-band, non-interacting picture of the GNR, and it would be worth checking how much of the oscillatory structure survives once a momentum-dependent scattering rate or electron-electron interactions are included, along the lines of the exact collision-integral treatments developed for other Landau-quantised systems~\cite{vavilov_aleiner_qbe, pongsangangan_ll_kinetic}. Tilt or anisotropy~\cite{ips-kush, ips-ruiz, ips_tilted_dirac, ips-sanskar, ips-kush-review, ips-tilted, yadav23_magneto}, not considered here, might distort the sawtooth and shift the resonance positions, offering a further handle for identifying a specific material realisation. Extending the present treatment to three-dimensional nodal-ring semimetals and their gapped counterparts, where the Fermi surface takes a toroidal shape~\cite{yang_review_nlsm, ips-nlsm-ph, ips-dipole-vnr, ips-gnr-strain}, is a natural next step, and could be compared directly against magnetotransport data on candidate compounds.

\section*{Acknowledgments}

We thank Muhammed Jaffar A. for participating in the initial stages of the project.

\appendix

\section{Starting formula for conductivity}
\label{appcond}

Within the LL-overlap approximation, the Green's functions are diagonal in the Landau-level basis, such that
\begin{equation}
G^{R}(E)=\sum_n |n\rangle\, G_n^{R}(E) \, \langle n| \,, \quad
G^{A}(E)=\sum_n |n\rangle \, G_n^{A}(E) \,\langle n|\,.
\end{equation}
Substituting this into Eq.~\eqref{eq:KB}, inserting the resolutions of the identity, $G^R(E)=\sum_m|m\rangle\,G_m^R(E)\,\langle m|$ and $G^A(E)=\sum_p|p\rangle\,G_p^A(E)\,\langle p|$, into $\mathrm{Tr}[v_\nu\,G^R(E)\,v_\mu\,G^A(E)]=\sum_n\langle n|\,v_\nu\,G^R(E)\,v_\mu\,G^A(E)\,|n\rangle$, and using $\langle p|n\rangle=\delta_{pn}$ to collapse the $p$-sum, we first keep $T>0$ general:
\begin{equation}
\mathrm{Tr}\!\left[v_\nu\,G^R(E)\,v_\mu\,G^A(E)\right]
=\sum_{n,m}\langle n|\,v_\nu\,|m\rangle\,\langle m|\,v_\mu\,|n\rangle\,G_m^R(E)\,G_n^A(E)\,,
\end{equation}
and Eq.~\eqref{eq:KB} therefore reads, at general temperature,
\begin{equation}
\label{eq:sig-finiteT}
\sigma_{\mu\nu}
=\frac{e^2}{2\,\pi\,\mathcal V}\int dE\left(  -\partial_E f\right)
\sum_{n,m}\langle n|\,v_\nu\,|m\rangle\,\langle m|\,v_\mu\,|n\rangle\,G_m^R(E)\,G_n^A(E)\,.
\end{equation}
The weight $-\partial_E f $ is the standard thermal broadening kernel,
\begin{equation}
- \partial_E f = \frac{\beta}{4}\,\mathrm{sech}^2\!\left(\frac{\beta(E-\upmu)}{2}\right)\,,
\quad \beta=1/T\,,
\end{equation}
representing a bump of unit weight [$\int dE\,(-\partial_E f)=f(-\infty)-f(\infty)=1$] centred at $E=\upmu$ with width $\sim T$. As $T\to0$ ($\beta\to\infty$) this bump narrows while its area stays fixed at unity, and it becomes a delta function,
\begin{equation}
\lim_{T\to0}\left(-\partial_E f \right) = \delta(E-\upmu)\,.
\end{equation}
Taking this $T\to0$ limit in Eq.~\eqref{eq:sig-finiteT}, the energy integral collapses onto $E=\upmu$, giving
\begin{equation}
\sigma_{\mu\nu}
=\frac{e^2}{2\,\pi\,\mathcal V}
\sum_{n,m}\mathrm{Re}\!\left[\langle n|\,v_\nu\,|m\rangle\,\langle m|\,v_\mu\,|n\rangle\,G_m^R(\upmu)\,G_n^A(\upmu)\right].
\end{equation}
Because $\sigma_{\mu\nu}$ denotes the dc conductivity, it is understood in the sense established after Eq.~\eqref{eq:KG}, with the physical dissipative response given by $\mathrm{Re}[\sigma_{\mu\nu}(\omega\to0)]$. The Kubo--Bastin expression in Eq.~\eqref{eq:KB} is already specialised to $\omega=0$, and the explicit $\mathrm{Re}[\cdots]$ simply reflects this convention. We also retain the Green's-function ordering $G_m^R(\mu)\,G_n^A(\mu)$ exactly as produced by the trace. The quantity $\sum_{n,m}\langle n|\,v_\nu \,|m\rangle
\langle m|\,v_\mu\, |n\rangle G_m^R \, G_n^A $
is already real once the double sum over $n$ and $m$ is performed: under $n\leftrightarrow m$, together with
$\langle n|v_\mu|m\rangle^*=\langle m|v_\mu|n\rangle $
and $G_n^A=(G_n^R)^*$, each term is mapped onto its complex conjugate. The explicit $\mathrm{Re}[\cdots]$ is retained only for consistency with the standard dc formulation.

Expressing the retarded and advanced Green's functions as
\begin{align}
& G_n^R=\frac{1}{D_n+i \, \Gamma_n}=\frac{1}{2}\left(\mathcal B_n-i\,\mathcal A_n\right),\quad
G_n^A=\frac{1}{D_n-i \, \Gamma_n}
=\frac{1}{2}\left(\mathcal B_n+i\, \mathcal A_n\right),
\quad
\mathcal A_n=\frac{2\,\Gamma_n}{D_n^2+\Gamma_n^2}\,,\quad
\mathcal B_n=\frac{2\,D_n}{D_n^2+\Gamma_n^2}\,,
\end{align}
leads to
\begin{align}
G_m^R \,G_n^A= \frac{1}{4}\left[
\mathcal A_n \,\mathcal A_m +\mathcal B_n\,\mathcal B_m
-i\left(\mathcal B_n\, \mathcal A_m
-\mathcal A_n\,\mathcal B_m \right)\right].
\end{align}
For the longitudinal conductivity, $ \langle n|\;v_x\;|m\rangle\langle m|\;v_x\;|n\rangle = |\langle n|\;v_x\;|m\rangle|^2 $ is purely real, and the real-part operation therefore retains only the real component of $ G_m^R \, G_n^A $, yielding the first expression in Eq.~\eqref{eq:sig-general}. For the Hall conductivity, Eq.~\eqref{eq:vyvx_matrix_element} shows that $ \langle n|\;v_y\;|m\rangle \, \langle m|\;v_x\;|n\rangle = i\,\mathrm{sgn}(m-n)\,|\langle n|\;v_x\;|m\rangle|^2$ is purely imaginary, and taking the real part of the product projects out the imaginary component of $G_m^R\,G_n^A$ instead. Summing over $n$ and $m$, the symmetric contribution proportional to $(\mathcal A_n \, \mathcal A_m + \mathcal B_n \, \mathcal B_m)$ cancels under $n\leftrightarrow m$, being itself symmetric while $\mathrm{sgn}(m-n)$ is antisymmetric, and what remains is the second expression in Eq.~\eqref{eq:sig-general}.

\section{The Fermi-sea contribution $\sigma_{xy}^{\rm II}$}
\label{app:sigmaII}

The Smr\v{c}ka--\v{S}treda decomposition splits the full linear-response Hall conductivity as $\sigma_{xy}=\sigma_{xy}^{\rm I}+\sigma_{xy}^{\rm II}$, where $\sigma_{xy}^{\rm I}$ is exactly the object computed via Eq.~\eqref{eq:KB} in the main text, and
\begin{align}
\label{eq:sigmaII-general}
\sigma_{xy}^{\rm II} = -\,\frac{e^2} {\mathcal V}
\sum_{E_\alpha<\upmu}\;\sum_{\beta\neq\alpha}
\frac{2\,\mathrm{Im}\!\left[\langle\alpha|\;v_x\;|\beta\rangle
\,\langle\beta|\;v_y\;|\alpha\rangle\right]} {(E_\alpha-E_\beta)^2}
\end{align}
is the clean-limit Kubo--TKNN sum over the filled Fermi sea, built entirely from retarded Green's functions, weighted by the full Fermi function rather than by $-\partial_E f$, and finite even as $\Gamma\to0$. For LL states $\alpha=(n,s)$, $\beta=(m,\tilde s)$, this sum diverges term by term because the hole branch $s=-$ is filled at every $n$. The inner sum runs over every $\beta\neq\alpha$ regardless of occupation, and this divergent piece is therefore identical at any $\upmu$ and at charge neutrality ($\upmu=0$). Subtracting the two cancels it exactly and leaves a finite sum over only the levels newly occupied between $\upmu=0$ and $\upmu$. For the Dirac cone this regularisation, combined with the \v{S}treda formula $\sigma_{xy}^{\rm II}=e\,\partial n/\partial B|_\upmu$, gives
\begin{align}
\sigma_{xy}^{\rm II}(\upmu) = \frac{e^2} {2 \, \pi}
\left[ N(\upmu)+\tfrac12 \right] \mathrm{sgn}(\upmu)\,,\quad N(\upmu)
=\left\lfloor\frac{\upmu^2}{\omega_B^2}\right\rfloor\,.
\end{align}
The half-integer shift traces back to the shared occupation of the zero-mode between the two branches.

For the GNR, the newly occupied conduction levels $n=0,\dots,N(\upmu)$ couple to both their same-branch and opposite-branch neighbours $m=n\pm1$. Going from the full state sum of Eq.~\eqref{eq:sigmaII-general} to a sum over LL indices $(n,m)$ alone means the $k_x$ sum has already been carried out; it contributes the same degeneracy factor $\mathcal V/(2\pi\ell_B^2)$ used throughout the main text, cancelling the $1/\mathcal V$ prefactor. The regularised sum accordingly retains an intraband and an interband piece,
\begin{align}
\sigma_{xy}^{\rm II}(\upmu) = -\,\frac{e^2\varepsilon_c}{2\,\pi\,m^*\,\ell_B^2}
\sum_{n=0}^{N(\upmu)}\sum_{m=n\pm1}
\mathrm{sgn}(m-n)\,\max(n,m) & \Bigg[
\frac{1}{(E_n-E_m)^2}
\left \lbrace 1-\frac{\Delta^2-(\varepsilon_n-\varepsilon_r)
\,(\varepsilon_m-\varepsilon_r)}{E_n \, E_m}\right \rbrace
\nn & +
\frac{1}{(E_n+E_m)^2}
\left \lbrace 1+\frac{\Delta^2-(\varepsilon_n-\varepsilon_r)
\,(\varepsilon_m-\varepsilon_r)}{E_n \, E_m}\right \rbrace \Bigg ]\,.
\end{align}
The interband term vanishes identically in the $\Delta\to0$, $\varepsilon_r\to0$ Dirac limit, recovering the intraband-only expression above, but remains generically nonzero for the GNR.

Rather than resumming this sum directly, its closed form follows more simply from \v{S}treda-formula level counting~\cite{smrcka}. The checkmark spectrum is non-monotonic, and $E_n<\upmu$ therefore admits two solutions in $n$ on the conduction branch, one on the descending side where $\varepsilon_n<\varepsilon_r$ and one on the ascending side where $\varepsilon_n>\varepsilon_r$. The filled conduction levels form a band between two field-dependent turning indices, $n_{\rm F}^{<}$ and $n_{\rm F}^{>}$, defined as in Ref.~\cite{barati-nlsm-qhe}. Counting this band gives
\begin{align}
\sigma_{xy}^{\rm II} = -\,\frac{e^2\,\mathcal C}{2 \, \pi}\,,\qquad \mathcal C = n_{\rm F}^{>}+1-n_{\rm F}^{<}\,.
\end{align}
Here, $n_{\mathrm F}^{<}$ and $n_{\mathrm F}^{>}$ are the highest-occupied LLs from the descending and ascending sets, respectively, and $\mathcal C$ is the Chern number of the filled particle-branch LLs as defined in Ref.~\cite{barati-nlsm-qhe}, whose result this appendix merely recasts in the regularised-sum language. As $B$ is varied at a fixed $\upmu$, $\mathcal C$ steps between consecutive integers each time a level crosses $\upmu$, producing the bumpy Hall plateaus noted therein, superposed on and independent of the disorder-broadened $\sigma_{xy}^{\rm I}$ computed in this paper.

\section{Nonzero matrix-elements for a Dirac cone}  
\label{appdirac}

For the Dirac cone, the velocity operators are $ v_x=v_F\,\sigma_x$ and $ v_y=v_F\,\sigma_y$, which act only on the pseudospin index of $|k_x,n,s\rangle$ and leave its plane-wave and oscillator factors untouched. The pseudospin matrix-elements are
\begin{align} 
\langle \tilde n,\tilde s|\,\sigma_x\,|n,s\rangle 
= \frac12 \,\Big( s\,\delta_{\tilde n,n+1} + \tilde s\,\delta_{\tilde n,n-1} \Big )\,,\quad
 \langle \tilde n,\tilde s|\,\sigma_y\,|n,s\rangle = \frac{i}{2}\,
  \Big ( -\, s\,\delta_{\tilde n,n+1} + \tilde s\,\delta_{\tilde n,n-1} \Big )\,. 
\end{align} 
Each matrix-element factorises exactly as in Eq.~\eqref{eqmat0} for the GNR: the plane-wave overlap $\frac{1}{L_x}\int dx\,e^{i \,(k_x-\tilde k_x)\,x}$ collapses onto $\delta_{k_x,\tilde k_x}$, which in turn pins the guiding centres of $\phi_{n-1}(y),\phi_n(y)$ and $\phi_{\tilde n-1}(y),\phi_{\tilde n}(y)$ to coincide. Thus, the residual oscillator overlaps reduce to plain orthonormality, $\int dy\,\phi_m(y)\,\phi_{m'}(y)=\delta_{mm'}$. Hence
\begin{align}
\langle \tilde k_x,\tilde n,\tilde s|\,v_x\,|k_x,n,s\rangle = \delta_{k_x,\tilde k_x}\,v_F\,
\langle \tilde n,\tilde s|\,\sigma_x\,|n,s\rangle\,,
\end{align}
and likewise for $v_y$. Therefore, 
\begin{align} 
|\langle k_x,n,s|\,v_x\,|\tilde k_x,\tilde n,\tilde s\rangle|^2
 \neq 0 \text{ iff }  |\tilde n-n|=1\,, 
\quad \langle k_x,n,s|\,v_y\,|\tilde k_x,\tilde n,\tilde s\rangle \langle \tilde k_x,\tilde n,\tilde s|\,v_x\,|k_x,n,s\rangle \neq 0 \text{ iff }  |\tilde n-n|=1\,. 
\end{align} 
Thus the Dirac case obeys the same vv selection rule as the GNR model.

In the ultraquantum limit, the chemical potential cuts the zeroth LL and expressions simplify considerably because there is only one zeroth LL. The vv selection rule requires $|\tilde n-n|=1$, and using $|k_x,0\rangle$ and $|k_x,n,s\rangle$ from Eqs.~\eqref{eq:dirac-eigenspinor} and~\eqref{eq:dirac-eigenspinor-LLL}, one finds 
\begin{align} 
& \langle \tilde k_x,0|\,v_x\,|k_x,n,s\rangle 
= \delta_{k_x,\tilde k_x}\,\frac{v_F}{\sqrt2}\,\delta_{n,1}\,, \quad
\langle \tilde k_x,0|\,v_y\,|k_x,n,s\rangle 
= \delta_{k_x,\tilde k_x}\,\frac{i\,v_F}{\sqrt2}\,\delta_{n,1}\,, 
\nn & \Rightarrow 
\big|\langle \tilde k_x,0|\,v_x\,|k_x,n,s\rangle\big|^2 
= \delta_{k_x,\tilde k_x}\,\frac{v_F^2}{2}\,\delta_{n,1}
\text{ and }  
\langle \tilde k_x,0|\,v_y\,|k_x,n,s\rangle \langle k_x,n,s|\,v_x\,|\tilde k_x,0\rangle 
= \delta_{k_x,\tilde k_x}\,\frac{i\,v_F^2}{2}\,\delta_{n,1} \,.
\end{align}
Summing over $k_x$ and $\tilde k_x$, the $k_x$ sum contributes the same Landau-degeneracy factor $\mathcal V/(2\pi\ell_B^2)$ [cf. Eq.~\eqref{eqdotsq}], leading to
\begin{align}
\sum_{k_x,\tilde k_x}\big|\langle \tilde k_x,0|\,v_x\,|k_x,n,s\rangle\big|^2
= \frac{\mathcal V}{2\,\pi\,\ell_B^2}\,\frac{v_F^2}{2}\,\delta_{n,1}
\end{align}
and
\begin{align}
& \sigma_{xx}^{ql}(\upmu)
=\sum_{s=\pm}\tilde\sigma_{xx}^{0;1,s}(\upmu)\,,\quad
\tilde\sigma_{xx}^{0;1,s}(\upmu)
=\frac{e^2\,v_F^2}{32\,\pi^2\,\ell_B^2}\,\mathcal A_{0}(\upmu)\,\mathcal A_{1,s}(\upmu)\,,
\nn &  \sigma_{xy}^{ql}(\upmu) = \sum_{s=\pm} \tilde\sigma_{xy}^{0;1,s}(\upmu)\,,
\quad \tilde\sigma_{xy}^{0;1,s}(\upmu) 
=\frac{e^2\,v_F^2}{16\,\pi^2\,\ell_B^2}
\left[ \mathcal B_{1,s}(\upmu)\, \mathcal A_{0}(\upmu) - \mathcal A_{1,s}(\upmu)
\, \mathcal B_{0}(\upmu) \right], \nn
& \mathcal{A}_{1,s}(\upmu) = 
\frac{2\,\Gamma_{1,s}} {(\upmu-s\,\omega_B)^2+\Gamma_{1,s}^{2}}\,, 
\quad \mathcal{A}_{0}(\upmu) = \frac{2\,\Gamma_{0}} {\upmu^2+\Gamma_{0}^{2}}\,,
\quad \mathcal{B}_{1,s}(\upmu) = \frac{2\,(\upmu-s\,\omega_B)}
 {(\upmu-s\,\omega_B)^2+\Gamma_{1,s}^{2}}\,, 
\quad \mathcal{B}_{0}(\upmu) = \frac{2\,\upmu} {\upmu^2+\Gamma_{0}^{2}}\,.
\end{align}
We note that $\mathcal V$ cancels exactly against the degeneracy factor, leaving a purely intensive result.
Since the LLL is pinned at zero energy for any $B$, the chemical potential in the ultraquantum limit sits exactly at $\upmu=0$, giving $\mathcal B_{0}(0)= 0$ and $\mathcal A_{0}(0) = 2/ \Gamma_0 $. Thus,
$$ \tilde\sigma_{xy}^{0;1,s}(0) = \frac{e^2\,v_F^2} {16\,\pi^2\,\ell_B^2}\,
\mathcal B_{1,s}(0)\,\mathcal A_{0}(0)
= \frac{e^2\,v_F^2}{16\,\pi^2\,\ell_B^2}\,
\frac{-\, 2\,s\,\omega_B}{\omega_B^2+\Gamma_{1,s}^{2}} \,\frac{2}{\Gamma_0}\,.$$
Because the two neighbouring levels $s=\pm$ are particle-hole symmetric about the zero of energy (since they sit at $\pm\,\omega_B$), their self-energies coincide: $\Gamma_{1,+}=\Gamma_{1,-}\equiv\Gamma_1$. The two terms in the sum over $s$ are then equal in magnitude and opposite in signs [viz. $\tilde\sigma_{xy}^{0;1,s}(0) \propto -\,s\,\omega_B$], yielding
\begin{align}
&  \sigma_{xx}^{ql}(0)= \frac{e^2 \, v_F^2 \, \Gamma_1} 
{4 \, \pi^2 \, \ell_B^2
\, \Gamma_0 \, (\omega_B^2+\Gamma_1^2)}\,,
\quad \omega_B^2= \frac{2 \, v_F^2 } {\ell_B^2}\,,
\quad \sigma_{xy}^{ql}(0) = \sum_{s=\pm}\tilde\sigma_{xy}^{0;1,s}(0) = 0\,,
\end{align}
independent of the impurity model. Although the disorder-broadened piece, $\sigma_{xy}^{\rm I}$, vanishes, the nonzero Fermi-sea contribution $\sigma_{xy}^{\rm II}$ (derived in Appendix~\ref{app:sigmaII}) evaluates to $\pm \,e^2/(4\,\pi)$ at $\upmu=0$ (with the sign fixed by which side of the zero mode $\upmu$ is pinned to). This is the familiar half-integer-shifted Hall sequence of a single Dirac cone \cite{Gusynin2005PRL, Zheng2002PRB}. The full Hall conductivity in the ultraquantum limit is therefore $\sigma_{xy}=\sigma_{xy}^{\rm I}+\sigma_{xy}^{\rm II}=\pm e^2/(4\pi)$, coming entirely from this topological Fermi-sea piece.

Unlike the LLL, whose self-energy $\Sigma_0$ is solved self-consistently since it is its own dominant intermediate state [cf.\ Eq.~\eqref{eqsel_LLL} for the GNR], the neighbouring level's self-energy $\Sigma_1$ must be evaluated non-self-consistently, using the already-determined $G_0$ rather than $G_1$ [cf.\ Eq.~\eqref{eqsel_neigh}]: $\Sigma_1(\upmu)=\vartheta_1/(\upmu-\Sigma_0(\upmu))$ and, thus, $\Sigma_1$ does not equal $\vartheta_1/(\upmu-\Sigma_1)$ (erroneously implemented in Ref.~\cite{dirac-qtm}). At $\upmu=0$ this gives $\Gamma_0=\sqrt{\vartheta_0}$ and $\Gamma_1=\vartheta_1/\Gamma_0=\vartheta_1/\sqrt{\vartheta_0}$ (rather than $\Gamma_1$ equalling $\sqrt{\vartheta_1}$). The 3 cases of disorder potentials are discussed below:
\begin{enumerate}

\item For pointlike disorder, the matrix-elements feeding into $\vartheta_0$ and $\vartheta_1$ are distinct: $\vartheta_0$ comes from the diagonal overlap $\langle 0|V|0\rangle \propto 
\int dy\,\phi_0^2 \, \phi_0^2$, while $\vartheta_1$ comes from the off-diagonal overlap $\langle 0|V|1,s\rangle\propto\int dy\,\phi_0\,\phi_1 \, \phi_0 \, \phi_1$ that dominates $\Sigma_1$ in the quantum limit. Evaluating both gives $\vartheta_1=\vartheta_0/2$ exactly (i.e.,  $\vartheta_1 \neq \vartheta_0$), so that $\Gamma_1=\Gamma_0/2$.

\item The Gaussian result follows the same substitution as above, with $\vartheta_0\propto1/S$, $\vartheta_1\propto1/S^2$, giving $\Gamma_1=\Gamma_0/(2S)$, where $S = 1+ 2\,R^2 /(\ell_B^2)$. This correctly reduces to the pointlike case at $R\to0$ (i.e., $S\to1$).

\item For the Yukawa case, we work in the same convention used in Sec.~\ref{sec:yukawa} for the GNR, i.e., with the general (not-yet RPA-fixed) screening wavevector $\kappa$ and the potential's amplitude $\tilde V_0^2=4\pi^2e^4/\epsilon^2$.\footnote{This is not the RPA-determined $\kappa$ of Ref.~\cite{dirac-qtm}, which we bring in below only for comparison.}
We specialise the general matrix-element of Eq.~\eqref{eq:yukawa-nn} to $n=0$ ($L_0^0=1$) and of Eq.~\eqref{eq:yukawa-adj} to $\lbrace n_<=0, \, n_>=1 \rbrace $ ($L_0^1=1$). Using the identities $\int_0^\infty\frac{e^{-u}}{u+\beta}\,du=e^\beta\Gamma_0(\beta)$ and $\int_0^\infty \frac{u\,e^{-u}}{u+\beta}\,du=1-\beta e^\beta \, \Gamma_0(\beta)$, we get
\begin{align}
\vartheta_0=M_{0,0}=\frac{n_{\rm imp}\,\tilde V_0^2}{4\,\pi}\,g\,e^{\mathcal K}\,,\quad
\vartheta_1=M_{0,1}=\frac{n_{\rm imp}\,\tilde V_0^2}{4\,\pi}
\left(1-\mathcal K\,g\,e^{\mathcal K}\right),
\quad \mathcal K= \frac{\kappa^2 \,\ell_B^2} {2} \,.
\end{align}
We note that $\vartheta_1>0$ for all $\mathcal K>0$, since ${\mathcal K} \,g\, e^{\mathcal K}<1$ identically (a standard property of $\Gamma_0=E_1$). Plugging in $\Gamma_0=\sqrt{\vartheta_0}$ and $\Gamma_1=\vartheta_1/\Gamma_0$ gives the Yukawa result.

\end{enumerate}
From the above discussion and via explicit computations, the final form of the longitudinal conductivity is captured by
\begin{align}
\label{eqdiracxx}
& \sigma_{xx}^{ql} = \dfrac{e^2}{16 \, \pi^2} \,\begin{cases} 
\left(1 + \dfrac{\gamma}{2}\right)^{-1} &\text{ for pointlike}\\
 \frac{1} {S} 
  \left [ 1 + \frac{\gamma}{2\,S^3} \right]^{-1} 
  &\text{ for Gaussian} \\
 \dfrac{2\left(1-{\mathcal K}\, g\,e^{\mathcal K}\right)}
 {g\,e^{\mathcal K}+\gamma_y\left(1-{\mathcal K}\, g\,e^{\mathcal K}\right)^2}
&\text{ for Yukawa}
\end{cases},\nn &
\gamma = \frac{ n_{\rm imp} \, {V_0}^2}{8 \, \pi \, v_F^2}\,, \quad 
S = 1+\frac{2\,R^2}{\ell_B^2}\,,\quad
\mathcal{K}= \frac{\ell_B^2\, \kappa^2} { 2}\,,\quad
g=\Gamma(\mathcal{K})\,,\quad
\gamma_y = \frac{n_{\rm imp}\,{\tilde V_0}^2\,\ell_B^2}{8\,\pi\,v_F^2}\,.
\end{align}
For the Yukawa case, we have defined the dimensionless disorder-strength $\gamma_y$ which carries an explicit $\ell_B^2\propto1/B$, because $\tilde V_0^2$ (unlike $V_0^2$) is not itself an area-normalised coupling. The result is expressed in terms of the free screening wavevector $\kappa$ (through $\mathcal K$).

It is worth comparing the ingredients feeding into $\sigma_{xx}^{ql}$ for the Dirac cone with that for the GNR, by looking closely at the Dirac spinors of Eqs.~\eqref{eq:dirac-eigenspinor}--\eqref{eq:dirac-eigenspinor-LLL} and the GNR spinor of Eq.~\eqref{eq:gnr-eigenspinor}. The result $\vartheta_1=\vartheta_0/2$ for the Dirac cone is caused by how the unit weight of the LLL's spinor is shared between the two spinorial components --- this is where the two models part ways. The Dirac zeroth level puts all of its weight in one component and leaves the other empty, while every other Dirac level splits its weight evenly between the two components. A disorder potential, being diagonal in pseudospin, connects only matching components, and this all-or-nothing versus even split is what fixes the clean factor of two between $\vartheta_0$ and $\vartheta_1$. The GNR spinor never collapses this way. Its two components are weighted by $A_n^{s\pm}=\sqrt{1\pm s\cos\theta_n}$ at every level, including whichever level plays the role of the LLL there, and this split depends continuously on $\rho$ and $\Delta$ through the mixing angle $\theta_n$. Carrying the same argument over to the GNR would therefore not give a universal factor of two between its own $\vartheta_0$ and $\vartheta_1$. That ratio there is instead a $\rho$- and $\Delta$-dependent quantity, set by the mixing angle rather than by a fixed spinor collapse. 

In the main text, we have treated the screening wavevector $\kappa$ as a free parameter for the Yukawa potential. In Ref.~\cite{dirac-qtm}, $\kappa$ is instead determined self-consistently via the random-phase approximation (RPA) for the static dielectric function of a massless Dirac cone in the ultraquantum limit. At zero temperature, the RPA screening wavevector is given by
\begin{align}
\kappa_{\rm rpa} = \frac{e^2}{2\,\pi\,\epsilon\,\ell_B v_F} \,,
\label{eq:yukawa-rpa}
\end{align}
where $\epsilon$ is the effective permittivity and $v_F$ is the Fermi velocity. This relation follows from the standard Lindhard function for a 2D Dirac system, evaluated at $\mathbf{q}=0$ with the chemical potential pinned at the zeroth LL.
Substituting $\kappa\to\kappa_{\rm rpa}$ into the Yukawa matrix-elements gives $\mathcal K \equiv \kappa^2\ell_B^2/2 = \alpha^2$, with $\alpha = e^2/(2\, \pi\, \epsilon \, v_F)$ the effective fine-structure constant of the system. Thus $\mathcal K$ becomes independent of the magnetic field. Consequently, $\vartheta_0=M_{0,0}$ and $\vartheta_1=M_{0,1}$ in Eq.~\eqref{eqdiracxx} are also $B$-independent (up to the constant $\ell_B^2$ that cancels with the $\ell_B^2$ in $\gamma_y$), and the only remaining field-dependence in $\sigma_{xx}^{ql}$ enters through $\omega_B^2 \propto 1/\ell_B^2$ in the denominator of Eq.~\eqref{eqdiracxx}.

Let us now figure out the asymptotic magnetic-field dependencies of Eq.~\eqref{eqdiracxx} in terms of the resistivity $\rho_{xx}^{ql}=1/\sigma_{xx}^{ql}$ (since $ \sigma_{xx}^{ql} = 0$):
\begin{enumerate}

\item For pointlike disorder, there is no $B$-dependence.

\item For the Gaussian potential, such limits follow from $ \rho_{xx}^{ql} \sim S\left [1 + {\gamma} / (2\,S^3) \right ]$,
where $S=1+u$, $u \equiv 2\, R^2/\ell_B^2$. For $u \gg 1$ ($S\approx u$), this gives $\rho_{xx}^{ql}\sim u\propto B$. For $u\ll 1$, expanding to first order gives $\rho_{xx}^{ql} \sim \left(1+\tfrac{\gamma}{2}\right) + u\,(1-\gamma) + \mathcal O(u^2)$, a constant plus a linear-in-$B$ correction (with a different coefficient than for the $ u \gg 1$ case, but the same qualitative linear-magnetoresistance behaviour).

\item For the Yukawa potential, holding $\kappa$ fixed (rather than using a RPA-determined value)
gives $\mathcal K=\ell_B^2 \,\kappa^2/2\propto B^{-1}$, leading to $\mathcal K\to0$ as $B\to\infty$. Since $g=\Gamma(\mathcal K)=E_1(\mathcal K)$ diverges logarithmically as $\mathcal K\to0$ (inferred from $g\sim-\ln\mathcal K-\gamma_E$), while $\mathcal K \,g \,e^{\mathcal K}\to 0$, $\vartheta_1 \to n_{\rm imp} \, \tilde V_0^2/(4 \, \pi)$ stays finite even as $\vartheta_0=n_{\rm imp} \,\tilde V_0^2 \,g \,e^{\mathcal K}/(4\,\pi)$ diverges. Thus, the denominator of $\sigma_{xx}^{ql}(0)$ is dominated by the $\omega_B^2 \,\vartheta_0$ term, giving $\sigma_{xx}^{ql} \sim e^2/(8\,\pi^2\, g)$. Furthermore, $g\sim-\ln\ell_B^2\sim\ln B$, implying $\rho_{xx}^{ql}\sim\ln B$. If instead $\kappa$ is fixed self-consistently through RPA screening as in Ref.~\cite{dirac-qtm}, $\mathcal K$ (and, hence, $\vartheta_0$ and $\vartheta_1$) is $B$-independent and only $\omega_B^2=2v_F^2/\ell_B^2$ varies. Expanding $\sigma_{xx}^{ql}$ in $\ell_B^2$ gives $\rho_{xx}^{ql}=\rho_\infty+c\,\ell_B^2+\mathcal O(\ell_B^4)$. Therefore,\ $\rho_{xx}^{ql}\sim B^0+\mathcal O(B^{-1})$, consistent with Ref.~\cite{dirac-qtm}'s description of the Yukawa case which reports a resistivity that approaches a constant from below with a leading $1/B$ correction. Thus the qualitative conclusion of Ref.~\cite{dirac-qtm} is unaffected and our corrected non-self-consistent treatment of the neighbouring level's self-energy only changes the numerical coefficients $\rho_\infty$ and $c$, not the functional form.

\end{enumerate}
Our asymptotic answers are summarised below:
\begin{align}
\rho_{xx}^{ql} \sim 
\begin{cases}
B^0 &\text{ for pointlike} \\[4pt]
B^{1} & \text{ for Gaussian} \; (R \gg \ell_B) \\[4pt]
B^0 + \mathcal O(B) & \text{ for Gaussian} \; (R \ll \ell_B)  \\[4pt]
\ln B & \text{ for Yukawa (generic fixed } \kappa \text{)} \\[4pt]
B^0 + \mathcal O(B^{-1})  & \text{ for Yukawa (RPA-fixed } \mathcal{K} \text{, as in Ref.~\cite{dirac-qtm})}
\end{cases}.
\label{eq:rho-scaling}
\end{align}

\bibliography{ref_nr_ll}

\end{document}